\RequirePackage{fix-cm}
\documentclass[twocolumn]{svjour3}          
\smartqed  
\usepackage[numbers]{natbib}
\usepackage{booktabs} 
\usepackage{graphicx}  
\usepackage{amssymb}
\usepackage{bm}
\usepackage{amsmath}
\usepackage{verbatim}
\usepackage[ruled,vlined]{algorithm2e}
\usepackage{subfigure}
\usepackage{upgreek} 
\usepackage{color}
\usepackage{ragged2e}

\usepackage{amsthm}
\usepackage{tabularx}
\usepackage{multirow}
\usepackage{verbatim}
\usepackage{url}            
\usepackage{tabularx}
\usepackage{multirow}
\usepackage{verbatim}

\begin{document}

\title{Visually Aware Recommendation with Aesthetic Features}


\author{Wenhui Yu \and Xiangnan He \and Jian Pei \and Xu Chen \and Li Xiong \and Jinfei Liu \and Zheng Qin}

\institute{Wenhui Yu \at
              Alibaba Group \\
              \email{jianlin.ywh@alibaba-inc.com}
           \and
           Xiangnan He \at
	           University of Science and Technology of China \\
	           \email{xiangnanhe@gmail.com}
	       \and
	       Jian Pei \at
	       Simon Fraser University \\
	       \email{jpei@cs.sfu.ca}
	       \and
	       Xu Chen (corresponding author) \at 
	       Beijing Key Laboratory of Big Data Management and Analysis Methods. Gaoling School of Artificial Intelligence, Renmin University of China, Beijing, China \\
	       \email{xu.chen@ruc.edu.cn}
	       \and
	       Li Xiong \at
	       Emory University \\
	       \email{lxiong@emory.edu}
	       \and
	       Jinfei Liu \at
	       Zhejiang University \\
	       \email{jinfeiliu@zju.edu.cn}
	       \and
	       Zheng Qin \at
	       Tsinghua Unversity \\
	       \email{qingzh@mail.tsinghua.edu.cn}
}

\date{Received: date / Accepted: date}

\maketitle

\begin{abstract}
Visual information plays a critical role in human decision-making process. Recent developments on visually aware recommender systems have taken the product image into account. We argue that the aesthetic factor is very important in modeling and predicting users' preferences, especially for some fashion-related domains like clothing and jewelry. This work is an extension of our previous paper \cite{Aes}, where we addressed the need of modeling aesthetic information in visually aware recommender systems. Technically speaking, we make three key contributions in leveraging deep aesthetic features. In \cite{Aes}, (1) we introduced the \textit{aesthetic features} extracted from product images by a deep aesthetic network to describe the aesthetics of products. We incorporated these features into recommender system to model users' preferences in the aesthetic aspect. (2) Since in clothing recommendation, time is very important for users to make decision, we designed a new tensor decomposition model for implicit feedback data. The aesthetic features were then injected to the basic tensor model to capture the temporal dynamics of aesthetic preferences. 

In this extended version, we try to explore aesthetic features in negative sampling to get further benefit in recommendation tasks. In implicit feedback data, we only have positive samples. Negative sampling is performed to get negative samples. In conventional sampling strategy, uninteracted items are selected as negative samples randomly. However, we may sample potential samples (preferred but unseen items) as negative ones by mistake. To address this gap, (3) we use the aesthetic features to optimize the sampling strategy. We enrich the pairwise training samples by considering the similarity among items in the aesthetic space (and also in the semantic space and graphs). The key idea is that a user may likely have similar perception on similar items. We perform extensive experiments on several real-world datasets and demonstrate the usefulness of aesthetic features and the effectiveness of our proposed methods.
\end{abstract}

\section{Introduction}
\label{sec:introduction}
Recommender systems have been widely used in online services to predict users' preferences based on their interaction histories \cite{NCF}. Recently, visual information has been intensively explored to enhance the performance of recommender models \cite{VBPR,Key_Frame,matrix,he_context}. In many domains of interest, the images of items play an important role in user decision-making process. For example, when purchasing clothing, users will scrutinize product images for the intuitive representation of the clothing like shape, design, color schemes, decorative pattern, texture, and so on. To leverage these kinds of information, existing efforts have extracted various visual features from item images and injected them into recommender models, like SIFT features, CNN features, color histograms, etc. For example, \citet{matrix,he_context} utilized low-level SIFT features and color histograms, and \citet{VBPR,Image_based,Key_Frame} utilized high-level CNN features extracted by a deep convolutional neural network.

\begin{figure}[ht!]
	\centering
	\includegraphics[scale = 0.38]{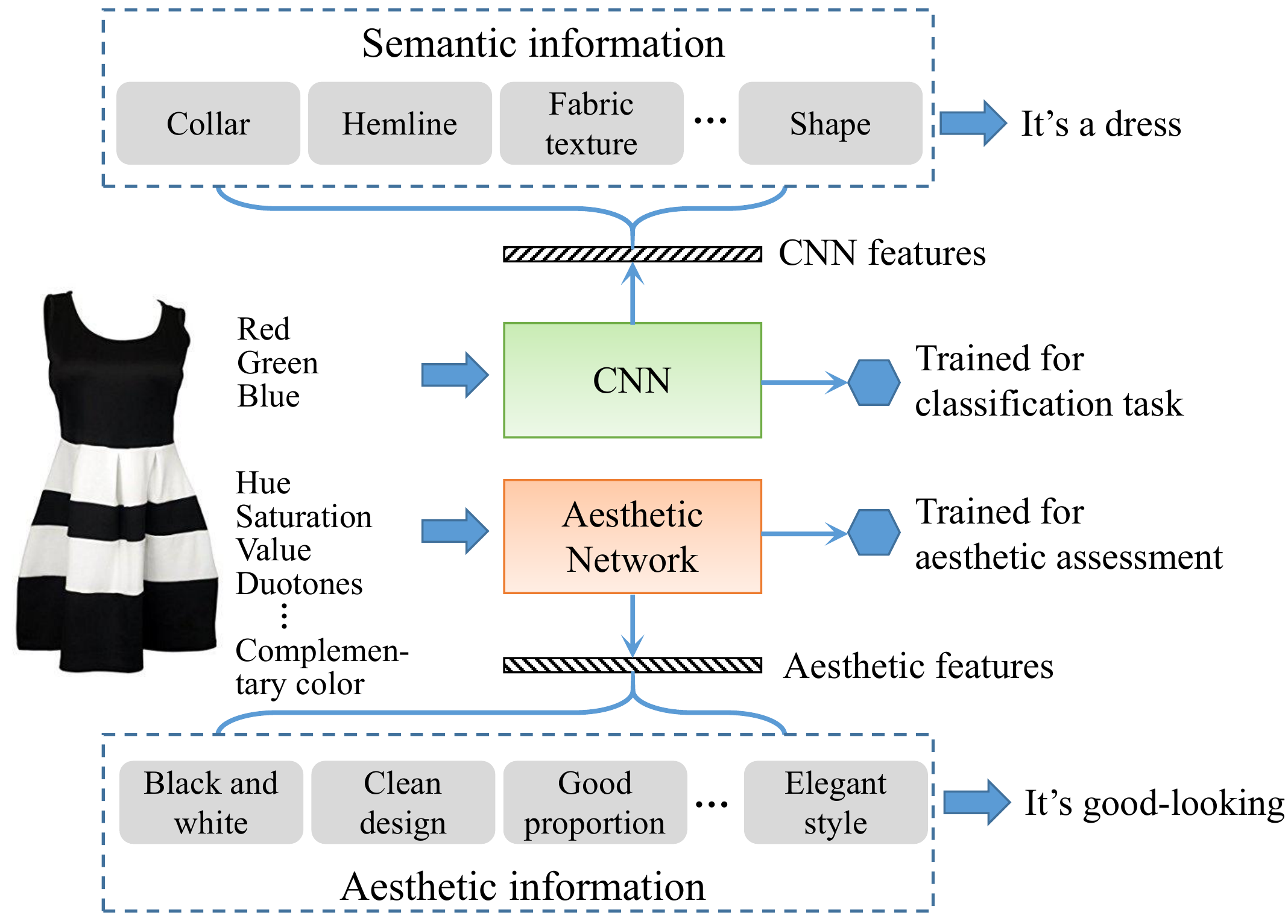}
	\caption{Comparison of CNN features and aesthetic features. The CNN is inputted with the RGB components of an image and trained for the classification task, while the aesthetic network is inputted with raw aesthetic features and trained for the aesthetic assessment task.}
	\label{fig:Comparison}
\end{figure}

We argue that the aesthetic information is crucial in predicting user preferences on products in many domains, such as clothing, furniture, ornaments, electronics, etc. Taking the product shown in Figure \ref{fig:Comparison} as an example, besides the semantic information, a user will also notice that the dress is with colors black and white, simple yet elegant design, and delightful proportion. She may have the intention to purchase it if she is satisfied with these aesthetic factors. In fact, for many users, especially young females, the aesthetic factor could be the primary factor when purchasing clothes. Unfortunately, conventional visual features do not encode the aesthetic information by nature. \citet{matrix} used color histograms to portray users' intuitive perception about an image, but the solution leaves much space to improve, since it does not make good use of many valuable information, such as aesthetic information shown in Figure \ref{fig:Comparison}. To address this issue, we proposed a more comprehensive and high-level aesthetic representation for items in our previous paper \cite{Aes}.

This paper is the extension of \cite{Aes}, where we extracted aesthetic-related features with a dedicated neural network called \textbf{B}rain-inspired \textbf{D}eep \textbf{N}etwork (\textbf{BDN}) \cite{Brain}. We input raw features that are indicative of human aesthetic feelings, such as hue, saturation, duotones, complementary colors, etc., and train BDN for the image aesthetic assessment task. We use the backbone to extract high-level aesthetic features. Intuitively, BDN is trained to mine information that is important to the aesthetic assessment task, thus these features encode aesthetic factors such as colors, structure, proportion, and styles (see Figure \ref{fig:Comparison} as an example). In this paper, to make a thorough use of aesthetic features, we additionally utilize them for negative sampling. In our method, the aesthetic features are used for both modeling and learning: we define the user preference model to be aware of aesthetic features, and then use them to improve the sampling quality when learning the model.

We first introduce the research effort in \cite{Aes}. Compared with other products, clothing shows obvious temporal characteristics, since in clothing recommendation, if an item can be purchased depends on not only if the user likes the it, but also if it fits the current time. To design the basic model, we consider these two vital factors. Also, users' aesthetic preferences are impacted by these two factors: (1) It is obvious that aesthetic preferences show a significant diversity among different people. For instance, when purchasing clothing, children prefer colorful and lovely products while adults prefer those can make them look mature and elegant (empirical evidence see Figure \ref{fig:s_age}); women may prefer exquisite decorations while men like concise designs (see Figure \ref{fig:v_gender}). (2) The aesthetic tastes of users also change over time, either in short term, or in long term. For example, the aesthetic tastes vary in different seasons periodically --- in spring or summer, people may prefer clothes with light color and fine texture, while in autumn or winter, people tend to buy clothes with dark color, rough texture, and loose style (see Figure \ref{fig:v_season}). In the long term, the fashion trend changes all the time and the popular colors and design may be different by year (see Figure \ref{fig:h_year}). 

Considering the above-mentioned factors, we exploit tensor factorization as the basic model to capture the diversity of aesthetic preferences among users and over time. There are several ways to decompose a tensor \cite{Tensor_application,Pairwise}, however, there are certain drawbacks in the existing models. To tailor it for the clothing recommendation task, we propose a new tensor factorization model trained with coupled matrices to mitigate the \emph{sparsity} problem \cite{All_at_once}. We then combine the basic model with the additional visual features (concatenated aesthetic and CNN features) and term the method \textbf{V}isually Aware \textbf{R}ecommendation with \textbf{A}esthetic Features (\textbf{VRA}).

Now, we introduce the research effort in this paper. The other technical contribution of the paper lies in the learning part. We not only leverage the features we extracted to model users' aesthetic preference in VRA, but also improve the quality of negative sampling by measuring the similarity of items in the aesthetic space. When optimizing a model on \textit{implicit feedback} data (e.g., purchasing records), pairwise learning has been widely used due to its rationality, which aims to maximize the margin between the predictions of positive and negative samples \cite{BPR}. In this paper, we design a pairwise learning to rank method to factorize the tensor and coupled matrices. However, when employing pairwise learning, one critical issue is that not all unobserved feedbacks are necessarily negative samples, since some of them might be just unknown by users, i.e., potential positive samples while mislabeled as negative ones. To address this issue, we construct the neighbor set of each item by finding the similar items in the aesthetic space. The intuition is that the items in the neighbor set of an purchased item (positive sample) are more likely to fit the user's aesthetics thus are more likely to be potential positive samples. We treat these potential positive samples as the third kinds of labels between the positive and negative ones in our \textbf{A}esthetic-enhanced \textbf{P}airwise \textbf{L}earning to \textbf{R}ank (\textbf{APLR}) algorithm.

Finally, we evaluate the performance of our proposed method by comparing it with several baselines on an \emph{Amazon} dataset and 5 subsets. Extensive experiments show that the recommendation accuracy can be significantly improved by incorporating aesthetic features. To summarize, our main contributions are as follows:

\begin{itemize}
	\item{We propose aesthetic features for items' aesthetic representation, and then leverage these features in the recommendation context. Moreover, we compare the effectiveness with several conventional features to demonstrate the necessity of the aesthetic features.}
	
	\item{We propose a new tensor factorization model to portray the purchase events in three dimensions: users, items, and time. We then inject the aesthetic features into it to model users' aesthetic preference.}
	
	\item{We use the aesthetic features to enhance the optimization strategy for the proposed model. To enrich pairwise training samples, we construct neighbor set for positive items by considering the similarity between items evidenced by visual features and collaborative information. This is the main contribution compared with our previous paper \cite{Aes}.}
	
	\item{We validate the effectiveness of our proposed model by comparing it against several state-of-the-art baselines on 6 real-world datasets. Experiments show that we gain significant improvement by exploring aesthetic features in modeling user preference and negative sampling.}
\end{itemize}

\section{Related Work}
\label{sec:related_work}
Recommender systems have gained more and more attention due to their extensive applications, and created considerable economic benefits. On various online platforms such as E-commerce, video, and news online platforms, recommender systems help users to find their interested items efficiently and improve the user experience significantly. Collaborative filtering (CF) model \cite{MF,item-based,user-based} boosts the development of recommender systems. Among various CF methods, matrix factorization (MF) \cite{MF,BPR}, which encodes user preferences by underlying latent factors, is a basic yet the most effective recommender model. To improve the presentation capability, many variants have been proposed \cite{VBPR,All_at_once,groupBPR,NCF,LCFN}.

This paper develops aesthetic-aware clothing recommender systems. Specifically, we incorporate the features extracted from the product images by an aesthetic network into a tensor factorization model, and optimize our model with pairwise learning. As such, we review related work on aesthetic networks, image-based recommendation, tensor factorization, and negative sampling strategies. 

\subsection{Aesthetic Networks}
The aesthetic networks are proposed for image aesthetic assessment. After \citet{Studying_aesthetic} first proposed the aesthetic assessment problem, many research efforts exploited various handcrafted features to extract the aesthetic information of images \cite{Studying_aesthetic,Content-based,Assessing_the_aesthetic_quality}. To portray the subjective and complex aesthetic perception, \citet{RAPID,Brain,O8} exploited deep networks to emulate the underlying complex neural mechanisms of human perception, and displayed the ability to describe image content from the primitive-level (low-level) features to the abstract-level (high-level) features. Proposed in \cite{Brain}, \textbf{B}rain-inspired \textbf{D}eep \textbf{N}etwork (\textbf{BDN}) model is the state-of-the-art aesthetic deep model. In this paper, we use BDN to extract the aesthetic features of product images, and use these features to enhance the performance of the recommender system.

\subsection{Image-based Recommendations}
Recommendation has been widely studied due to its extensive use. The power of recommender systems lies on their ability to model complex preferences that users exhibit toward items based on their past interactions and behavior. To extend their expressive power, various works exploited image data \cite{VBPR,Key_Frame,matrix,he_Attentive,he_context}. For example, \citet{Image_based,VBPR} used CNN features of product images while \citet{matrix} recommended movies with color histograms of posters and frames. \citet{clothes1,clothes2} recommended clothes by considering the clothing fashion style. Though various visual features are leveraged in recommendation tasks, they are conventional features (such as CNN features and SIFT features) and low-level aesthetic features (such as color histograms). To propose more powerful aesthetic features, \citet{Aes} extracted high-level features by a BDN pretrained for the aesthetic assessment task, and used these features to model users' aesthetic preference. This paper is the extended version of \cite{Aes}. To explore aesthetic features in different aspects, we used them to improve the quality of negative sampling.

\subsection{Tensor Factorization}
Time is an important contextual information in recommender systems since the sales of commodities show a distinct time-related succession. In context-aware recommender systems, tensor factorization has been extensively used. For example, \citet{Tensor_application} introduced two main forms of tensor decomposition, the \textbf{C}ANDECOMP/ \textbf{P}ARAFAC (\textbf{CP}) and Tucker decomposition. \citet{O10} first utilized tensor factorization for context-aware collaborative filtering. \citet{Pairwise} proposed a \textbf{P}airwise \textbf{I}nteraction \textbf{T}ensor \textbf{F}actorization (\textbf{PITF}) model to decompose the tensor with a linear complexity. Tensor-based methods suffer from several drawbacks like poor convergence in sparse data \cite{poor_convergence} and not scalable to large-scale datasets \cite{Scalable}. To address these limitations, \citet{All_at_once,who} formulated recommendation models with the \textbf{C}oupled \textbf{M}atrix and \textbf{T}ensor \textbf{F}actorization (\textbf{CMTF}) framework. All existing tensor decomposition models are designed for explicit feedback data and usually do not perform well in implicit feedback cases. In this paper, we design a novel tensor decomposition model for implicit feedback data and incorporate aesthetic features into it.

\subsection{Negative Sampling}
In real-world applications, data of implicit feedback, or \textit{one-class} form is easier to collect so extensively used. Prediction on implicit feedback dataset is a challenging task since we only know positive samples and unobserved samples, but cannot discriminate negative samples and potential positive samples from the unobserved ones \cite{O2}. In \cite{BPR}, all unobserved samples are treated equally as negative ones when sampling. To improve the sampling quality, many works proposed enhanced pairwise learning with various extra information \cite{viewBPR,adaptiveBPR,dynamicBPR,groupBPR,itemgroupBPR}. For example, \citet{viewBPR,adaptiveBPR} used view information to enrich positive samples. \citet{groupBPR,CPLR} utilized collaborative information mined from the connections of users and items. \citet{listrank,Listwise} proposed listwise ranking methods instead of pairwise ones. \citet{NBPO} considered the noise in negative samples, and optimized the negative sampling strategy based on noisy label-robust learning. \citet{TDAR} performed negative supervision on the embedding level by domain adaptation, thus can avoid negative sampling. 

Though widely explored, the effectiveness of high-level visual features in this task is neglected. In this paper, we leverage the aesthetic features (additionally with semantic features and collaborative information) in the learning to rank process. For each positive item, we regard items with similar visual features or items connected in the bipartite graph as the neighbors (potential positive samples), and assume that users will prefer them to other negative samples.

\begin{figure*}[ht!]
	\centering
	\includegraphics[scale = 0.5]{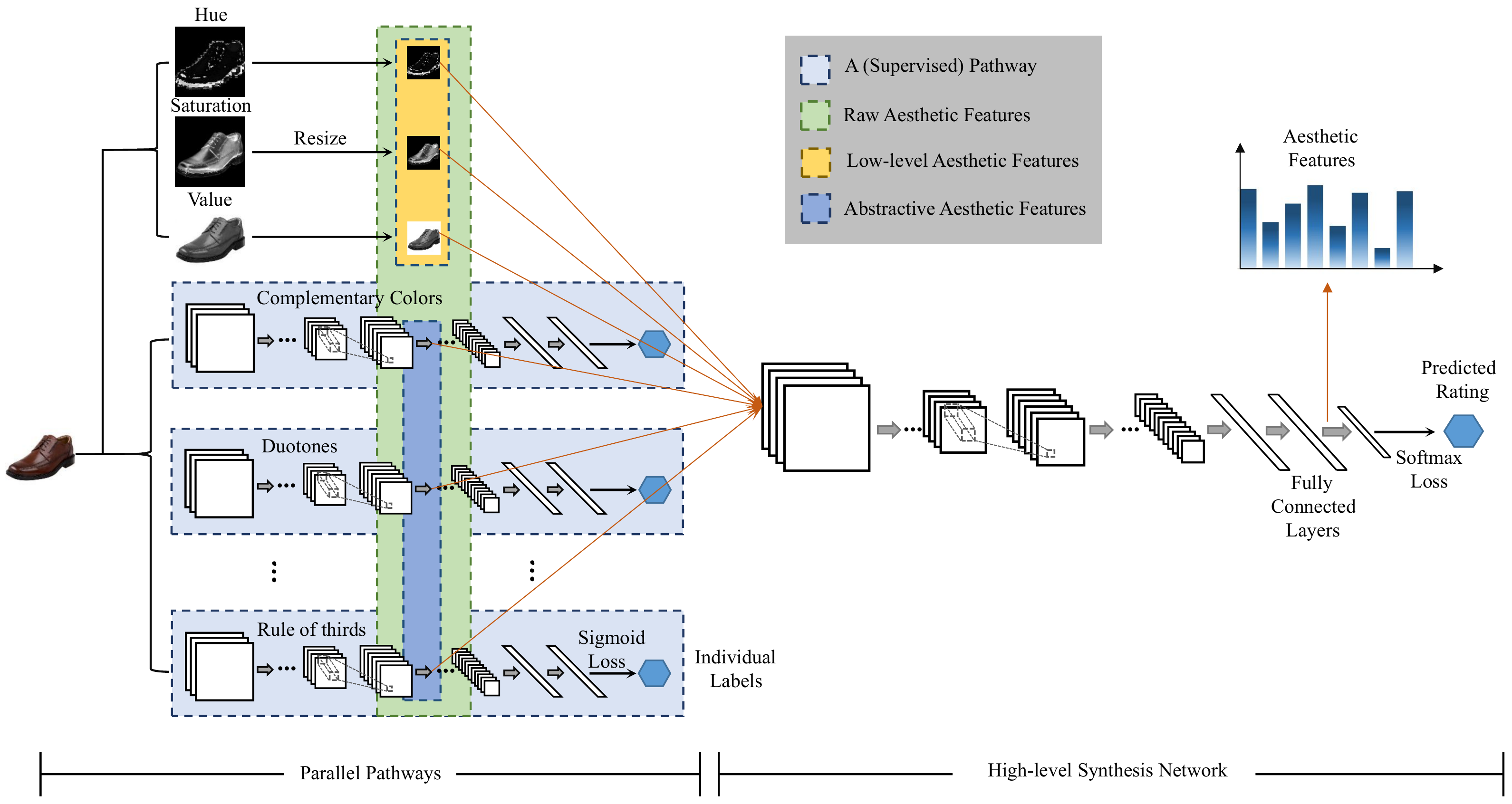}
	\caption{Brain-inspired Deep Network (BDN) architecture.}
	\label{fig:aesthetic}
\end{figure*}

\section{Preliminaries}
\label{sec:preliminaries}
In this section, we introduce some preliminaries about the aesthetic neural network, which is used to extract the aesthetic features of clothing images. 

\citet{Brain} introduced the Brain-inspired Deep Networks (BDN, shown in Figure \ref{fig:aesthetic}), a deep CNN structure consists of several parallel pathways (sub-networks) and a high-level synthesis network. It is trained on the \emph{Aesthetic Visual Analysis (AVA)} dataset, which contains 250,000 images with aesthetic ratings and tagged with 14 photographic styles (e.g., complementary colors, duotones, rule of thirds, etc.). The pathways take the form of convolutional networks to exact the abstractive aesthetic features by \emph{pre-trained} with the individual labels of each tag. For example, when training the pathway for complementary colors, the individual label is 1 if the sample is tagged with ``complementary colors'' and is 0 if not. 

We input the raw features, including low-level features (hue, saturation, value) and abstractive features (feature maps of the pathways), into the high-level synthesis network and \emph{jointly tune} it with the pathways for aesthetic rating prediction. Considering that the \emph{AVA} is a photography dataset and the styles are for photography, so not all the raw features extracted by the pathways are desired in our recommendation task, thus we only reserve the pathways that are relevant to the clothing aesthetic. Finally, we use the output of the second fully-connected layer of the synthesis network as our aesthetic features.

We then analyze several extensively used features to illustrate the superiority of our aesthetic features.

\textbf{CNN Features:} These are the most extensively used features due to their extraordinary representation ability. Trained for the image classification task, CNN extracts the features important to image semantics, thus CNN features mainly contain semantic information, which contributes little to evaluate the aesthetics of an image. Recall the example in Figure \ref{fig:Comparison}, it can encode ``There is a skirt in the image'' but cannot express ``The clothing is beautiful and fits the user's taste''. Devised for aesthetic assessment, BDN can capture the high-level aesthetic information. As such, our aesthetic features can do better in beauty estimating and complement CNN features in clothing recommendation.

\textbf{Color Histograms:} \citet{matrix} exploited color histograms to represent human's feeling about the posters and frames for movie recommendation. Though can get the aesthetic information roughly, the low-level handcrafted features are crude, unilateral, and empirical. BDN can get abundant visual features by the pathways. Also, it is data-driven, since the rules to extract features are learned from the data. Compared with the intuitive color histograms, our aesthetic features are more objective and comprehensive. Recall the example in Figure \ref{fig:Comparison} again, color histograms can tell us no more than ``The clothes in the image is white and black''.

\section{Aesthetic-based Recommendation}
\label{sec:aesthetic_model}
In this section, we first introduce the basic tensor factorization model, and then integrate visual features into the basic model to propose the Visually Aware Recommendation with Aesthetic Features (VRA) model. The summary of notations are represented in Table~\ref{tab:notation}.

\begin{table*}[ht!]
	\caption{The summary of notations.}
	\centering
	\label{tab:notation}
	\scalebox{1}{
		\begin{tabular}{cc}
			\toprule
			Notations & Definitions\\
			\midrule
			$p$/$q$/$r$ & the $p$-th user/$q$-th item/$r$-th time \\
			$P$/$Q$/$R$ & the total number of users/items/time intervals \\
			${\bm{{\rm A}}}$/${\bm{{\rm B}}}$/${\bm{{\rm C}}}$ & user-item-time tensor/user-item matrix/time-item matrix\\
			$\hat{\bm{{\rm A}}}$/$\hat{\bm{{\rm B}}}$/$\hat{\bm{{\rm C}}}$ & reconstruction of ${\bm{{\rm A}}}$/${\bm{{\rm B}}}$/${\bm{{\rm C}}}$ \\
			${\bm{{\rm F}}}$/${\bm{{\rm f}}}_{CNN}$/${\bm{{\rm f}}}_{AES}$ & visual features/CNN features/aesthetic features \\
			${\bm{{\rm \Theta}}}$ = \{${\bm{{\rm U}}}$, ${\bm{{\rm V}}}$, ${\bm{{\rm W}}}$, ${\bm{{\rm T}}}$, ${\bm{{\rm M}}}$, ${\bm{{\rm N}}}$\} & model parameters \\
			$\lambda_c$/$\lambda_r$ & the weighting parameter/regularization coefficient\\
			\hline
			$\mathcal{P}$ & the set of $P$ users $\{p_1,p_2, ...,p_P \}$\\
			$\mathcal{Q}$ & the set of $Q$ items $\{q_1,q_2, ...,q_Q \}$\\
			$\mathcal{R}$ & the set of $R$ items $\{r_1,r_2, ...,r_R \}$\\
			$\mathcal{Q}^+_p$/$\mathcal{Q}^-_p$ & the set of items purchased/not purchased by user $p$\\
			$\mathcal{Q}^+_r$/$\mathcal{Q}^-_r$ & the set of items purchased/not purchased in time $r$\\
			$\mathcal{Q}^+_{pr}$/$\mathcal{Q}^-_{pr}$ & the set of items purchased/not purchased by user $p$ and/or in time $r$\\
			$\mathcal{D}$ & the training set of (user, positive item, time) tuples\\
			$\mathcal{N}_q^C$ & the neighbour set of item $q$ constructed based on CNN features\\
			$\mathcal{N}_q^A$ & the neighbour set of item $q$ constructed based on aesthetic features\\
			$\mathcal{N}_q^U$ & the neighbour set of item $q$ constructed based on CNN users \\
			$\mathcal{N}_q^T$ & the neighbour set of item $q$ constructed based on time\\
			$\mathcal{A} \setminus \mathcal{B}$ & the set of elements in $\mathcal{A}$ but not in $\mathcal{B}$\\
			$|\mathcal{A}|$ & the size of set $\mathcal{A}$\\
			\bottomrule
	\end{tabular}}
\end{table*}

\subsection{Basic Model}
\label{subsec:dcf}
Considering the impact of time on aesthetic preferences, we propose a context-aware model as the basic model to account for the temporal factor. We use a $P \times Q \times R$ tensor ${\bm{{\rm A}}}$ to indicate the purchase events among the user, clothes, and time dimensions (where $P$, $Q$, $R$ are the number of users, clothes, and time intervals, respectively). If user $p$ purchased item $q$ in time interval $r$, ${\bm{{\rm A}}}_{pqr}=1$, otherwise ${\bm{{\rm A}}}_{pqr}=0$. Tensor factorization has been widely used to predict the missing entries (i.e., zero elements) in ${\bm{{\rm A}}}$, which can be used for recommendation. 

\subsubsection{Existing Methods and Their Limitations} In this subsection, we summarize the motivation of our novel tensor factorization model by revealing the limitations of existing models.

\textbf{Tucker Decomposition:} This method \cite{Tensor_application} decomposes the tensor ${\bm{{\rm A}}}$ into a tensor core and three matrices, 
$$\hat{\bm{{\rm A}}}_{pqr}=\sum_{i=1}^{K_1} \sum_{j=1}^{K_2} \sum_{k=1}^{K_3} {\bm{{\rm a}}}_{ijk} {\bm{{\rm U}}}_{ip} {\bm{{\rm V}}}_{jq} {\bm{{\rm T}}}_{kr},$$
where ${\bm{{\rm a}}} \in \mathbb{R}^{K_1 \times K_2 \times K_3}$ is the tensor core, ${\bm{{\rm U}}} \in \mathbb{R}^{K_1 \times P}$, ${\bm{{\rm V}}} \in \mathbb{R}^{K_2 \times Q}$, and ${\bm{{\rm T}}} \in \mathbb{R}^{K_3 \times R}$. Tucker decomposition has very strong representation ability, but it is very time consuming, and hard to converge.

\textbf{CP Decomposition:} The tensor ${\bm{{\rm A}}}$ is decomposed into three matrices in CP decomposition,
$$\hat{\bm{{\rm A}}}_{pqr}=\sum_{k=1}^{K} {\bm{{\rm U}}}_{kp} {\bm{{\rm V}}}_{kq} {\bm{{\rm T}}}_{kr},$$
where ${\bm{{\rm U}}} \in \mathbb{R}^{K \times P}$, ${\bm{{\rm V}}} \in \mathbb{R}^{K \times Q}$, and ${\bm{{\rm T}}} \in \mathbb{R}^{K \times R}$. This model has been widely used due to its linear time complexity, especially in Coupled Matrix and Tensor Factorization (CMTF) structure models \cite{All_at_once,who,Scalable}. However, all dimensions (users, clothes, time) are mapped to the same latent factor space. Intuitively, we want the latent factors relating users and clothes to encode the information about users' preference, like aesthetics, prices, quality, brands, etc., and the latent factors relating clothes and time to encode the information about the seasonal characteristics and fashion elements of clothes like colors, thickness, design, etc. 

\textbf{PITF Decomposition:} The Pairwise Interaction Tensor Factorization (PITF) model \cite{Pairwise} decomposes ${\bm{{\rm A}}}$ into three pair of matrices,
$$\hat{\bm{{\rm A}}}_{pqr}=\sum_{k=1}^{K} {\bm{{\rm U}}}_{kp}^{\bm{{\rm V}}} {\bm{{\rm V}}}_{kq}^{\bm{{\rm U}}} +\sum_{k=1}^{K} {\bm{{\rm U}}}_{kp}^{\bm{{\rm T}}} {\bm{{\rm T}}}_{kr}^{\bm{{\rm U}}} + \sum_{k=1}^{K} {\bm{{\rm V}}}_{kq}^{\bm{{\rm T}}} {\bm{{\rm T}}}_{kr}^{\bm{{\rm V}}},$$
where ${\bm{{\rm U}}}^{\bm{{\rm V}}}, {\bm{{\rm U}}}^{\bm{{\rm T}}} \in \mathbb{R}^{K \times P}$; ${\bm{{\rm V}}}^{\bm{{\rm U}}}, {\bm{{\rm V}}}^{\bm{{\rm T}}} \in \mathbb{R}^{K \times Q}$; ${\bm{{\rm T}}}^{\bm{{\rm U}}}, {\bm{{\rm T}}}^{\bm{{\rm V}}} \in \mathbb{R}^{K \times R}$. PIFT has a linear complexity and strong representation ability. Yet, it is not in line with implicit feedbacks due to the additive combination of each pair of matrices. For example, in PIFT, for certain clothes $q$ liked by the user $p$ but not fitting the current time $r$, $q$ gets a high score for $p$ and a low score for $r$. It should not be recommended to the user since we want to recommend the right item in the right time. However, the total score can be high enough if $p$ likes $q$ so much that $q$'s score for $p$ is very high. In this case, $q$ will be returned even it does not fit the time. In addition, PITF model is inappropriate to be trained with coupled matrices.

\subsubsection{Model Formulation}
To address the limitations of the aforementioned models, we propose a new tensor factorization method which is for implicit feedback with linear complexity. When a user makes a purchase decision on a clothing product, there are two primary factors: if the product fits the user's preferences and if it fits the time. A clothing product fits a user's preferences if the appearance is appealing, the style fits the user's tastes, the quality is good, and the price is acceptable. And a clothing product fits the time if it is in-season and fashionable. For user $p$, clothing $q$, and time interval $r$, we use the scores $S_1$ and $S_2$ to indicate how the user likes the clothing and how the clothing fits the time respectively. $S_1 = 1$ when the user likes the clothing and $S_1 = 0$ otherwise. Similarly, $S_2 = 1$ if the clothing fits the time and $S_2 = 0$ otherwise. The user will buy the clothing only if $S_1 = 1$ and $S_2 = 1$, so, $\hat{{\bm{{\rm A}}}}_{pqr} = S_1 \& S_2$. To make the formula differentiable, we can approximately formulate it as $\hat{{\bm{{\rm A}}}}_{pqr} = S_1 \cdot S_2$. We present $S_1$ and $S_2$ in the form of matrix factorization: $S_1 = {\bm{{\rm U}}}_{*p}^\mathsf{T} {\bm{{\rm V}}}_{*q}$, $S_2 = {\bm{{\rm T}}}_{*r}^\mathsf{T} {\bm{{\rm W}}}_{*q}$, where ${\bm{{\rm U}}} \in \mathbb{R}^{K_1 \times P}$, ${\bm{{\rm V}}} \in \mathbb{R}^{K_1 \times Q}$, ${\bm{{\rm T}}} \in \mathbb{R}^{K_2 \times R}$, and ${\bm{{\rm W}}} \in \mathbb{R}^{K_2 \times Q}$. The prediction is then given by:
\begin{eqnarray}
\label{equ:newpair}
\hat{{\bm{{\rm A}}}}_{pqr} = \left({\bm{{\rm U}}}_{*p}^\mathsf{T} {\bm{{\rm V}}}_{*q}\right) \left({\bm{{\rm T}}}_{*r}^\mathsf{T} {\bm{{\rm W}}}_{*q}\right).
\end{eqnarray}
We can see that in Equation (\ref{equ:newpair}), the latent factors relating users and clothes are independent with those relating clothes and time. Though the $K_1$-dimensional vector ${\bm{{\rm V}}}_{*q}$ and the $K_2$-dimensional vector ${\bm{{\rm W}}}_{*q}$ are all latent factors of clothing $q$, ${\bm{{\rm V}}}_{*q}$ captures the information about users' preferences whereas ${\bm{{\rm W}}}_{*q}$ captures the temporal information of the clothing. Compared with CP decomposition, our model is more effective and expressive in capturing the underlying latent patterns in purchases. Compared with PITF, combining $S_1$ and $S_2$ with \& (approximated by multiplication) is helpful to recommend right clothing in right time. Moreover, our model is efficient and easy to train compared with the Tucker decomposition.

\subsubsection{Coupled Matrix and Tensor Factorization}
Though widely used to portray the context information in recommendation, tensor factorization suffers from poor convergence due to the sparsity of the tensor. To relieve this problem, \citet{All_at_once} proposed a CMTF model, which decomposes the tensor with coupled matrices. In this subsection, we couple our tensor factorization model with restrained matrices during training. As our model is proposed by considering two factors: $S_1$ (use's preference towards items) and $S_2$ (time's ``preference'' towards items), we also explore restrained matrices that can supervise these two factors.

\textbf{User $\times$ Clothing Matrix:} We use matrix ${\bm{{\rm B}}} \in \mathbb{R}^{P \times Q}$ to indicate the purchase activities between users and clothes. ${\bm{{\rm B}}}_{pq} = 1$ if user $p$ purchased clothing $q$ and ${\bm{{\rm B}}}_{pq} = 0$ if not. We use ${\bm{{\rm B}}}$ to supervise $S_1$ when learning our model.

\textbf{Time $\times$ Clothing Matrix:} We use matrix ${\bm{{\rm C}}} \in \mathbb{R}^{R \times Q}$ to record when the clothing was purchased. Since the characteristics of clothing change steadily with time, we make a coarse-grained discretization on time to avoid the tensor from being extremely sparse. Time is divided into $R$ intervals in total. ${\bm{{\rm C}}}_{rq} = 1$ if clothing $q$ is purchased in time interval $r$ and ${\bm{{\rm C}}}_{rq} = 0$ if not. We use ${\bm{{\rm C}}}$ to supervise $S_2$.

In previous work \cite{All_at_once,who,O10,O11}, the CMTF models are optimized by minimizing the reconstruction loss ($\rm MSE\_O\scriptstyle PT$):
\begin{small}
\begin{flalign}
\label{equ:mse}
\rm MSE\_O\scriptstyle PT = &\frac{1}{2} {\left\Arrowvert \bm{{\rm A}} - \hat{\bm{{\rm A}}} \right\Arrowvert}_{\rm F}^2
+ \frac{\lambda_c}{2} \left({\left\Arrowvert \bm{{\rm B}} - \hat{\bm{{\rm B}}} \right\Arrowvert}_{\rm F}^2
+ {\left\Arrowvert \bm{{\rm C}} - \hat{\bm{{\rm C}}} \right\Arrowvert}_{\rm F}^2\right) \nonumber\\
& + \frac{\lambda_r}{2} {\left\Arrowvert \bm{{\rm \Theta}} \right\Arrowvert}_{\rm F}^2
\qquad
\end{flalign}
\end{small}where $\hat{\bm{{\rm A}}}$, $\hat{\bm{{\rm B}}}$, and $\hat{\bm{{\rm C}}}$ are the reconstructions of ${\bm{{\rm A}}}$, ${\bm{{\rm B}}}$, and ${\bm{{\rm C}}}$, respectively. $\hat{\bm{{\rm A}}}$ is defined in Equation (\ref{equ:newpair}), $\hat{\bm{{\rm B}}} = {\bm{{\rm U}}}^\mathsf{T} {\bm{{\rm V}}}$, and $\hat{\bm{{\rm C}}} = {\bm{{\rm T}}}^\mathsf{T} {\bm{{\rm W}}}$; $\lambda_c$ is a parameter to balance the weights of the tensor term and coupled matrix terms. The last term of Equation (\ref{equ:mse}) is the regularization term to prevent overfitting, and $\lambda_r$ is the regularization coefficient. ${\left\Arrowvert \;\; \right\Arrowvert}_{\rm F}$ is the Frobenius norm of a matrix, ${\bm{{\rm \Theta}}}$ represents the parameters of the model, ${\bm{{\rm \Theta}}} = \{{\bm{{\rm U}}}, {\bm{{\rm V}}}, {\bm{{\rm T}}}, {\bm{{\rm W}}}\}$. As shown in Equation (\ref{equ:mse}), we train model parameters to complete ${\bm{{\rm A}}}$, and use ${\bm{{\rm B}}}$ and ${\bm{{\rm C}}}$ to assist the supervision of model training. 

\begin{figure}[ht!]
	\centering
	\subfigure[Item examples]{
		\includegraphics[scale = 0.45]{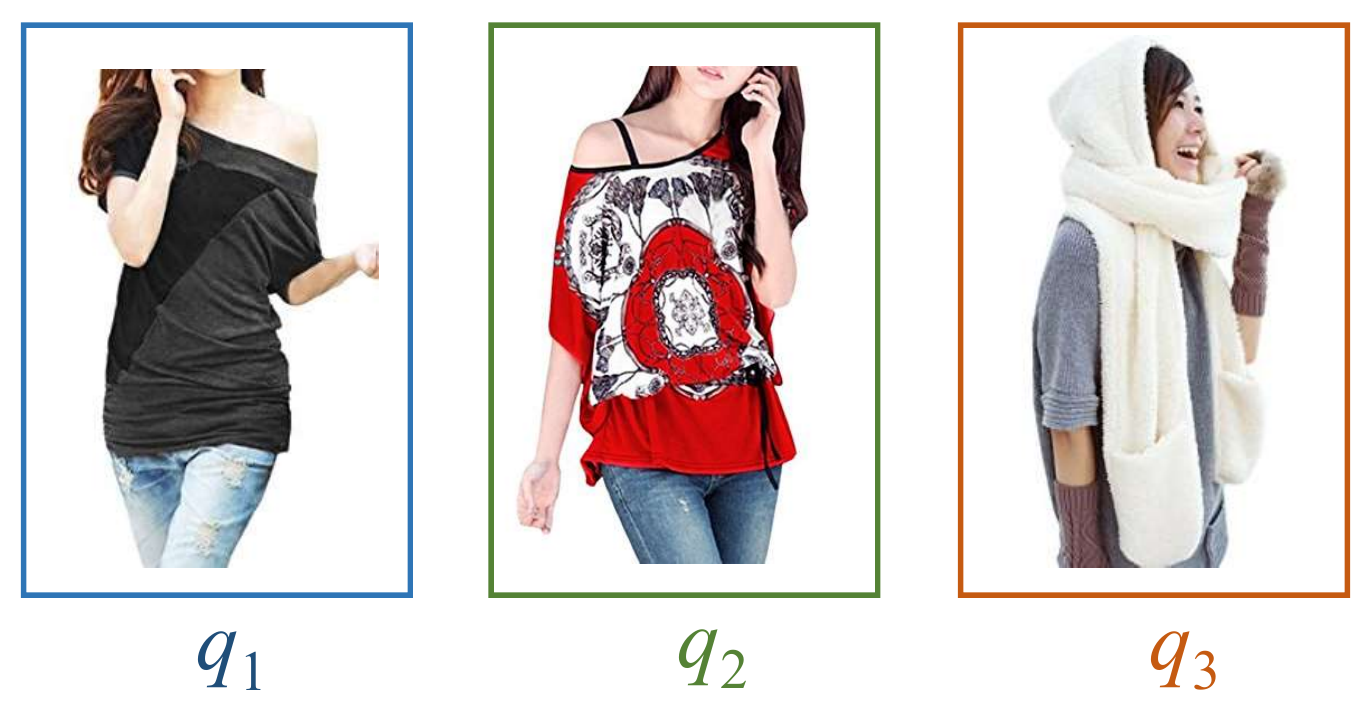}
		\label{fig:exam1}
	}\\
	\subfigure[User-item latent space]{
		\includegraphics[scale = 0.45]{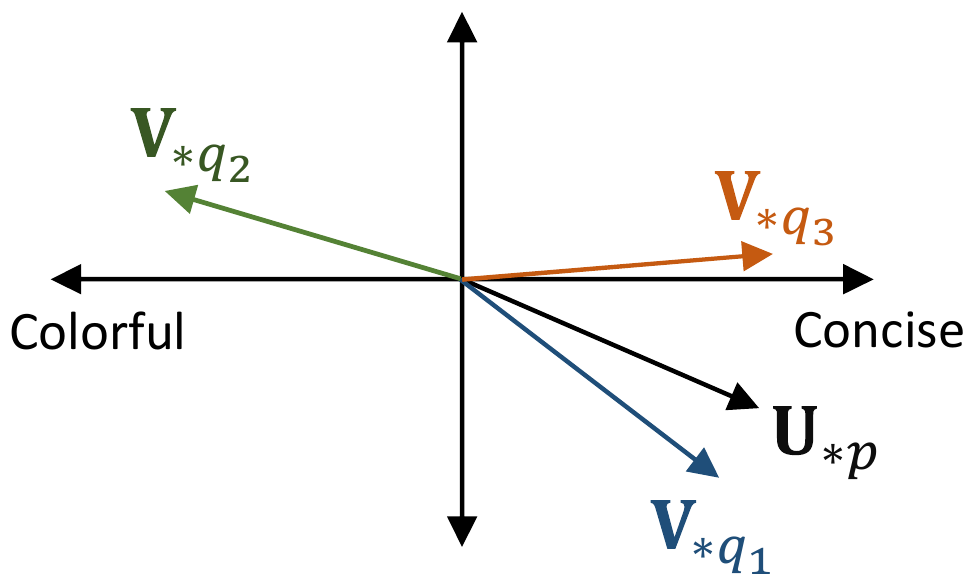}
		\label{fig:exam2}
	}
	\subfigure[Time-item latent space]{
		\includegraphics[scale = 0.45]{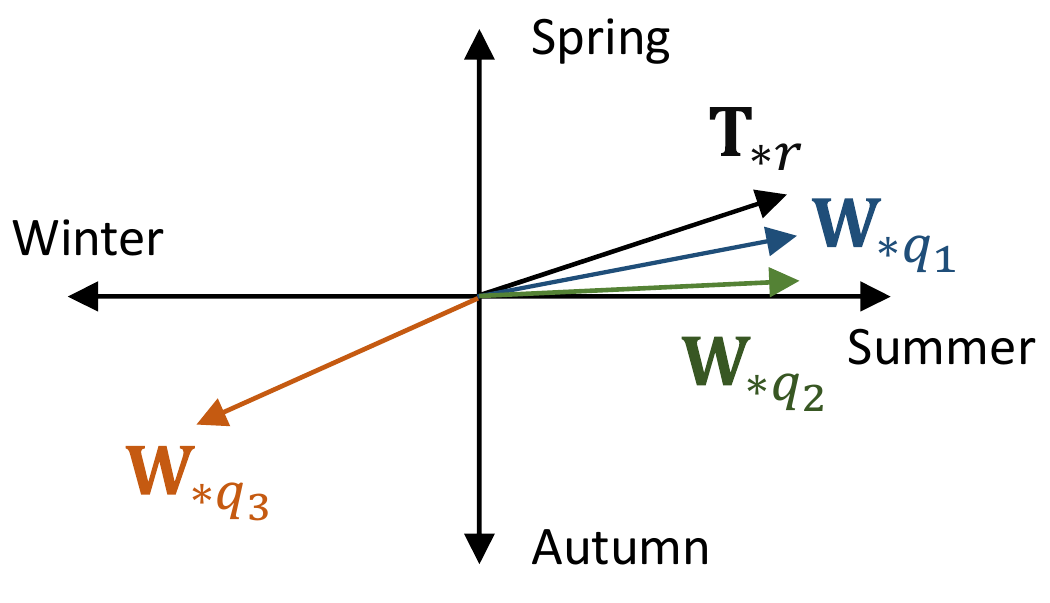}
		\label{fig:exam3}
	}
	\caption{An example to illustrate our basic model.}
	\label{fig:example}
\end{figure}

\newtheorem{myEXP}{Example}
\begin{myEXP}
	\label{exp:basic_model}
	We give an example to illustrate how our basic model works (please see Figure~\ref{fig:example}). There are three items ($q_1$, $q_2$, and $q_3$) and two latent factor spaces (the user-item latent space and time-item latent space). The user-item latent space encodes the users' preference and the time-item latent space encodes the temporal characteristics of items. In our basic model, we map users and items into user-item latent space by ${\bm{{\rm U}}}$ and ${\bm{{\rm V}}}$, and map time intervals and items into time-item latent space by ${\bm{{\rm T}}}$ and ${\bm{{\rm W}}}$. In this example, we aim to recommend clothes to a user $p$ who likes simple and elegant clothes in a time interval $r$ in summer. For clothing $q_1$, we can see that it fits $p$'s preference and it is a shirt designed for summer, thus $q_1$ gets high $S_1$ and $S_2$ scores and can be recommended due to the high score $S = S_1\cdot S_2$. For the clothing $q_2$, it is a piece of summer clothes yet is too colorful for $p$, thus $q_2$ gets low $S_1$ score and high $S_2$ score and cannot be recommended. Clothing $q_3$ is simple and elegant yet is used in winter, thus $q_3$ gets high $S_1$ score and low $S_2$ score and cannot be recommended either.
	
	{\color{black} If we neglect the first term in Equation (\ref{equ:mse}), predicting $S_1$ in the user-item latent space and predicting $S_2$ in the time-item space are two independent recommendation tasks. Supervised by ${\bm{{\rm B}}}$, predicting $S_1$ is a conventional recommendation task which recommends items to users. Supervised by ${\bm{{\rm C}}}$, predicting $S_2$ is to ``recommend'' items to current time. When predicting $S_2$, we need to encode ``preferences'' of time in the time-item latent space. These ``preferences'' may relate to the seasonal or fashion information.}
\end{myEXP}

\subsection{Hybrid Model}

In this section, we incorporate the visual features into the basic model, and optimize it with the pairwise learning to rank method.

\subsubsection{Model Formulation}

Combined with visual features, we formulate the predictive model as:
\begin{eqnarray}
\label{equ:pair_image}
\hat{{\bm{{\rm A}}}}_{pqr} = \left({\bm{{\rm U}}}_{*p}^\mathsf{T} {\bm{{\rm V}}}_{*q} + {\bm{{\rm M}}}_{*p}^\mathsf{T} {\bm{{\rm F}}}_{*q}\right) \left({\bm{{\rm T}}}_{*r}^\mathsf{T} {\bm{{\rm W}}}_{*q} + {\bm{{\rm N}}}_{*r}^\mathsf{T} {\bm{{\rm F}}}_{*q}\right),
\end{eqnarray}where ${\bm{{\rm F}}} \in \mathbb{R}^{2K \times Q}$ is the feature matrix, ${\bm{{\rm F}}}_{*q}$ is the visual features of clothing $q$, which is the concatenation of CNN features (${\bm{{\rm f}}}_{CNN}\in \mathbb{R}^{K \times 1}$) and aesthetic features (${\bm{{\rm f}}}_{AES}\in \mathbb{R}^{K \times 1}$), ${\bm{{\rm F}}}_{*q} = \left[ \begin{matrix} {\bm{{\rm f}}}_{CNN}\\{\bm{{\rm f}}}_{AES} \end{matrix} \right]$ and $K = 4096$. ${\bm{{\rm M}}} \in \mathbb{R}^{2K \times P}$ and ${\bm{{\rm N}}} \in \mathbb{R}^{2K \times R}$ are visual preference matrices. ${\bm{{\rm M}}}_{*p}$ encodes the visual preferences of user $p$ and ${\bm{{\rm N}}}_{*r}$ encodes the visual preferences in time interval $r$. In our model, both the latent factors and visual features contribute to the final prediction. Though the latent factors can uncover any relevant attributes theoretically, they usually cannot in real-world applications on account of the sparsity of the data and lack of information. So the assistance of visual information can highly enhance the model. Also, recommender systems often suffer from the \emph{cold start} problem. We cannot extract information for users and clothes without consumption records in CF methods. In this case, extra (visual and context) information can alleviate this problem. For example, for certain ``cold'' clothing $q$, we can decide whether to recommend it to a certain user $p$ in current time $r$ according to if $q$ looks satisfying to the user (determined by ${\bm{{\rm M}}}_{*p}$) and to the time (determined by ${\bm{{\rm N}}}_{*r}$).

\begin{figure*}[ht!]
	\centering
	\includegraphics[scale = 0.45]{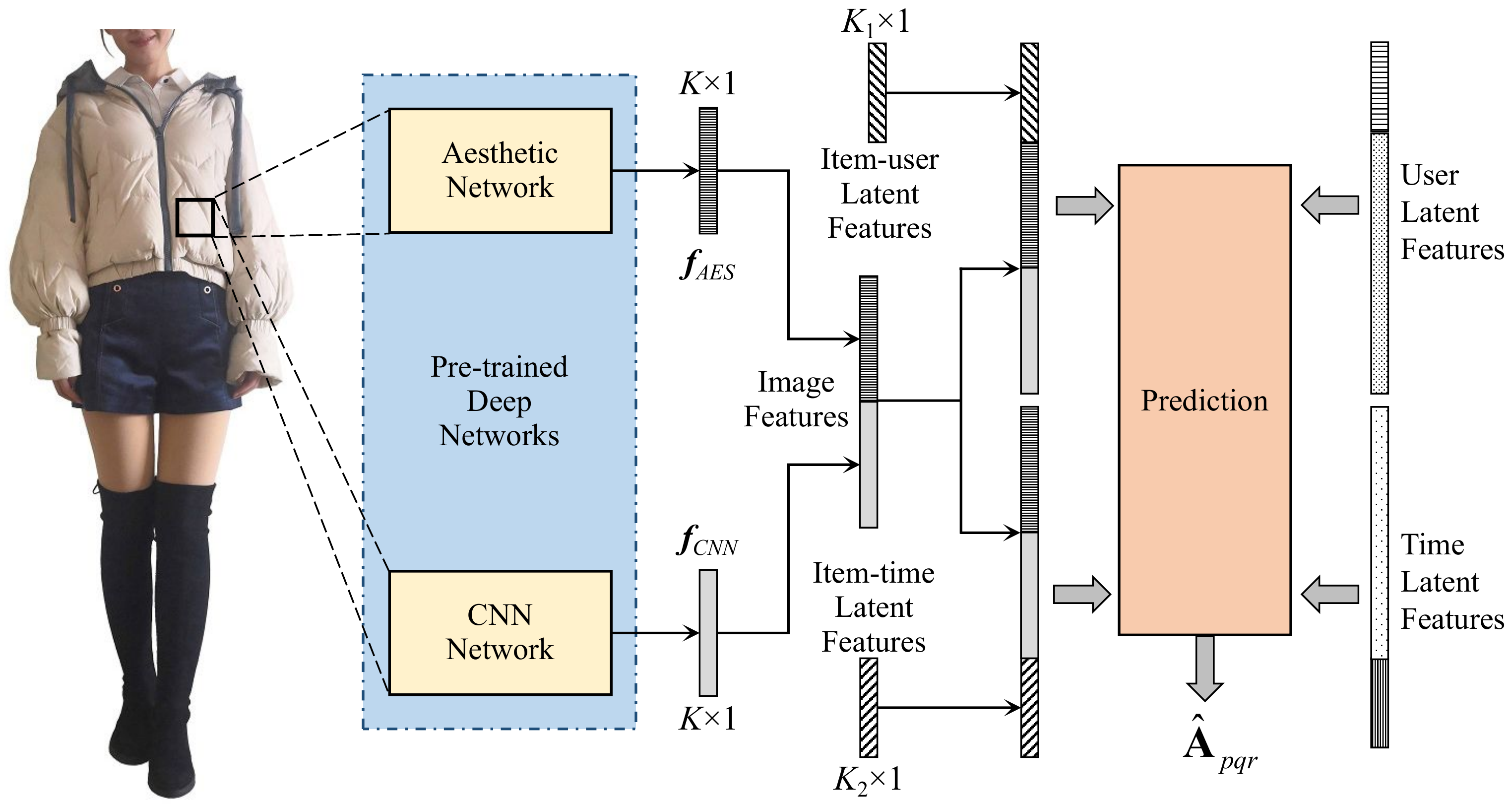}
	\caption{Diagram of our preference predictor.}
	\label{fig:predictor}
\end{figure*}

\subsubsection{Pairwise Learning to Rank}
\label{sec:PLR}
Since the Mean Squared Error Optimization ($\rm MSE\_O\scriptstyle PT$, please see Equation (\ref{equ:mse})), which is widely used in existing CMTF models \cite{All_at_once,who,O10,O11}, is designed for explicit feedback data, we design \textbf{P}airwise \textbf{L}earning to \textbf{R}ank (\textbf{PLR}) method with coupled matrix constrain for our VRA on implicit feedback data. We represent the positive set $\mathcal{D}$ in the form of triples: $$\mathcal{D} = \big\{(p,q,r)\big| {\bm{{\rm A}}}_{pqr}=1 \big\},$$ and the set of unlabeled samples is: $$\mathcal{Q}_{pr}^- = \big\{q\big|q\in \mathcal{Q} \setminus \big(\mathcal{Q}^+_{p} \cup \mathcal{Q}^+_{r}\big)\big\},$$ where $\mathcal{Q}$ denotes the set of items, $\mathcal{Q}^+_{p} = \big\{q\big| {\bm{{\rm B}}}_{pq}=1 \big\}$ denotes the set of items purchased by user $p$, and $\mathcal{Q}^+_{r} = \big\{q\big| {\bm{{\rm C}}}_{rq}=1 \big\}$ denotes the set of items purchased in time $r$. The objective function is formulated as:
\begin{small}
	\begin{flalign}
	\label{equ:objective_function}
	{\rm PLR\_O{\scriptstyle PT}} = \sum_{(p,q,r)\in \mathcal{D}} \sum_{q'\in \mathcal{Q}_{pr}^-} L \left(p,q,q',r \right) - \frac{\lambda_r}{2} {\left\Arrowvert {\bm{{\rm \Theta}}} \right\Arrowvert}_{\rm F}^2\, .
	\end{flalign}
\end{small}
$L(\,\,)$ in Equation (\ref{equ:objective_function}) is the likelihood function, \begin{small}
	$$L\left (p,q,q',r\right )={\rm ln} \, \sigma \left (\hat{\bm{{\rm A}}}_{pqq'r}\right ) + \lambda_c \Big[ {\rm ln} \, \sigma \left (\hat{\bm{{\rm B}}}_{pqq'}\right ) + {\rm ln} \, \sigma \left (\hat{\bm{{\rm C}}}_{rqq'}\right ) \Big],$$
\end{small}where $\hat{\bm{{\rm A}}}$ is defined in Equation (\ref{equ:pair_image}), $\hat{\bm{{\rm B}}} = {\bm{{\rm U}}}^\mathsf{T} {\bm{{\rm V}}} + {\bm{{\rm M}}}^\mathsf{T} {\bm{{\rm F}}}$, and $\hat{\bm{{\rm C}}} = {\bm{{\rm T}}}^\mathsf{T} {\bm{{\rm W}}} + {\bm{{\rm N}}}^\mathsf{T} {\bm{{\rm F}}}$; $\hat{\bm{{\rm A}}}_{pqq'r} = \hat{\bm{{\rm A}}}_{pqr} - \hat{\bm{{\rm A}}}_{pq'r}$, $\hat{\bm{{\rm B}}}_{pqq'} = \hat{\bm{{\rm B}}}_{pq} - \hat{\bm{{\rm B}}}_{pq'}$, $\hat{\bm{{\rm C}}}_{rqq'} = \hat{\bm{{\rm C}}}_{rq} - \hat{\bm{{\rm C}}}_{rq'}$; $\sigma( \,\,\,)$ is the sigmoid function; The model is optimized from users' \emph{implicit feedback} with mini-batch gradient descent, which calculates the gradient with a small batch of samples.

\section{Aesthetic-enhanced Pairwise Learning to Rank}
\label{sec:pairwise_learning}
In Section \ref{sec:aesthetic_model}, we leverage aesthetic features to model users' aesthetic preference and in this section, we use aesthetic features to improve ranking performance (CNN features and collaborative information are leveraged as well for comprehensiveness). PLR which is introduced in the Subsection \ref{sec:PLR} is a pairwise learning method for multi-objective optimization with the aim of maximizing the gap between the positive feedbacks and negative feedbacks. Pairwise learning has been widely used due to its strong performance \cite{he_Attentive,VBPR,HLBPR} while there is a critical issue in the current formulation. To be specific, a user did not purchase a product may because she is not interested in it, but may also because that she has never seen it before. However, in pairwise learning, all missing entries are treated as negative samples hence many potential positive samples are mislabeled as negative ones. To to uncover these potential positive samples, we construct the neighbor set $\mathcal{N}_q$ for each positive sample $q$ by uncovering the products that have similar visual representations with $q$, or the products connected to $q$ in the user-item or time-item graphs. In other words, $\mathcal{N}_q$ contains the products near $q$ in the visual space or in the graphs. If a user purchased $q$, she may also prefer $\mathcal{N}_q$ due to the similarity on comprehensive aspects. In this section, we propose a Aesthetic-enhanced Pairwise Learning to Rank (APLR) by considering these potential positive samples in ranking.

\subsection{Problem Formulation}
\label{subsec:item_collaborative}
When sampling, we regard the neighbours as potential positive samples. For a user $p$ and a time interval $r$, we assume that (1) user $p$ prefers items with positive feedbacks to the others; (2) user $p$ prefers the neighbours of the positive sample to the irrelevant ones; (3) positive samples fit the current time $r$ better than the others; (4) neighbours of the positive sample fit the current time $r$ better than the irrelevant ones. So for each $(p,q,r)$ in $\mathcal{D}$, we have the preference relationship,
\begin{flalign}
&(p,q,r)\succ(p,\mathcal{Q}_{pr}^-,r), (p,q,r)\succ(p,\mathcal{N}_q,r),\nonumber\\ 
&(p,\mathcal{N}_q,r)\succ(p,\mathcal{Q}_{pr}^-\setminus\mathcal{N}_q,r).\nonumber
\end{flalign}
As such, we can generalize Equation (\ref{equ:objective_function}) as follows:
\begin{small}
	\begin{flalign}
	\label{equ:objective_function2}
	{\rm APLR\_O{\scriptstyle PT}}&\!\! =\!\!\!\!\! \sum_{(p,q,r)\in \mathcal{D}}\!\! \Bigg[ \!\sum_{q''\in \mathcal{Q}_{pr}^-} \!\!\!\!L\left(p,q,q''\!,r\right) \!+\!\eta_1\!\!\! \sum_{q'\in \mathcal{N}_q}\!\!\! L\left(p,q,q'\!,r\right) \nonumber\\
	&+\eta_2\!\! \sum_{q'\in \mathcal{N}_q} \sum_{q''\in \mathcal{Q}_{pr}^-\setminus \mathcal{N}_q} \!\!\!\!L\left(p,q',q'',r\right)\Bigg]
	\!- \!\frac{\lambda_r}{2} {\left\Arrowvert {\bm{{\rm \Theta}}} \right\Arrowvert}_{\rm F}^2 , 
	\end{flalign}
\end{small}where $\eta_1$ and $\eta_2$ are weighting parameters. Here we can see that for each purchase record $(p,q,r)$, user $p$ prefers $q$ to $q'$ and prefers $q'$ to $q''$. The preference relationship is constructed by finding the neighbours of the positive items, which can be interpreted as an item-based collaborative learning model \cite{NAIS}. Most existing works learn to rank by constructing the potential set of each user \cite{CPLR,groupBPR,socialBPR,viewBPR,itemgroupBPR,adaptiveBPR}. In the next subsections, we will introduce how to construct neighbour set for each item, and demonstrate the advantages of our item-based collaborative sampling strategy.

\subsection{Constructing Neighbor Set}
\label{subsec:nei_set}

To find the neighbors of each positive sample, we leverage the visual information and the collaborative information. For visual information, we cluster all products with CNN features and aesthetic features. For each product, the cluster it belongs to is the neighborhood set. And for collaborative information, we find all products purchased by the same user or purchased in the same time to be the neighbor products.

\textbf{Neighbors in aesthetic space:} Similarly, we cluster all products by the aesthetic features and regard the cluster a product $q$ belongs to as the aesthetic neighbor set, denoted as $\mathcal{N}_q^A$. Products close to each other in the aesthetic space have similar aesthetic characteristics. For a certain user, since that positive samples are in line with her aesthetics, neighbors are also in line with her aesthetics.

\textbf{Neighbors in semantic space:} We cluster all products by the CNN features. For a product $q$, the cluster it belongs to is the semantic neighbor set, denoted as $\mathcal{N}_q^C$. Products with similar CNN features have similar appearances, users may have interests in the items that look like the purchased ones.

\textbf{Neighbors linked by users:} For each product $q$, we find all products that purchased by the same user to consist the user-linked neighbor set, $\mathcal{N}_q^U = \big\{q'\big\vert {\bm{{\rm B}}}_{pq}=1 \wedge {\bm{{\rm B}}}_{pq'}=1\big\}$. Each product $q'$ in $\mathcal{N}_q^U$ has been purchased by the same user with $q$, therefore users who have interests in $q$ may also like $q'$. We update the part of our model which captures the users' preferences (parameters ${\bm{{\rm U}}}$, ${\bm{{\rm V}}}$, and ${\bm{{\rm M}}}$) with $\mathcal{N}_q^U$.

\textbf{Neighbors linked by time:} For each product $q$, we find all products that purchased in the same time with $q$ to consist the time-linked neighbor set, $\mathcal{N}_q^T = \big\{q'\big\vert {\bm{{\rm C}}}_{rq}=1 \wedge  {\bm{{\rm C}}}_{rq'}=1\big\}$. Each product $q'$ in $\mathcal{N}_q^T$ has been purchased in the same time with the current product $q$, so $q'$ may fit the current time better than other missing value samples. We update the part which captures the temporal character of products in our model (parameters ${\bm{{\rm T}}}$, ${\bm{{\rm W}}}$, and ${\bm{{\rm N}}}$) with $\mathcal{N}_q^T$.

$\mathcal{N}_q^U$ is the neighbour set of $q$ in user-item bipartite graph and $\mathcal{N}_q^T$ is that in time-item bipartite graph. Of special notice is that they are used to update different parts of our model. Taking $\mathcal{N}_q^T$ as an example, it only contributes to predicting $S_2$. As we discussed in Example~\ref{exp:basic_model}, when predicting $S_2$, we recommend items to each time interval, i.e., we capture the ``preference'' of current time $r$ rather than of current user $p$, and return ``personalized'' recommendation to $r$. In this situation, two items $q_1$ and $q_2$ that both fit $r$ are two similar items from time perspective, though they may be totally different when considering the preference of the user, therefore we only use $\mathcal{N}_q^T$ to update $\{{\bm{{\rm T}}},{\bm{{\rm W}}},{\bm{{\rm N}}}\}$ rather than $\{{\bm{{\rm U}}},{\bm{{\rm V}}},{\bm{{\rm M}}}\}$.
	
$\mathcal{N}_q^T$ is an extension of $\mathcal{N}_q^U$ from the user-item graph to time-item graph. Considering the difference between user $p$ and time $r$ ($p$ is an index and $r$ is a discrete numerical value), a more general way to construct $\mathcal{N}_q^T$ is to set a window $\Delta r$, and for each product $q$, $\mathcal{N}_q^T$ contains all products that purchased in similar time (in range of $r \pm \Delta r$) with $q$, i.e. $\mathcal{N}_q^T = \big\{q'\big\vert {\bm{{\rm C}}}_{rq}=1 \wedge {\bm{{\rm C}}}_{r'q'}=1 \wedge |r-r'|\leq \Delta r\big\}$. Since the density of ${\bm{{\rm C}}}$ is much higher than ${\bm{{\rm B}}}$, the size of $\mathcal{N}_q^T$ will be very large when we set a large $\Delta r$, we set $\Delta r = 0$ in our APLR.

\subsection{Model Learning}

We then calculate the gradient of Equation (\ref{equ:objective_function2}). To maximize the objective function, we take the first-order derivatives with respect to each model parameter:
\begin{small}
	\begin{flalign}
	\label{derivatives}
	{\nabla_{\!\bm{{\rm \Theta}}}{\rm APLR\_O\scriptstyle PT}} &\!=\!\!\!\!\!\!\!\sum_{(p,q,r)\in \mathcal{D}}\!\! \Bigg[ \sum_{q''\in \mathcal{Q}_{pr}^-} \!\!\!\!\!\frac{\partial L\!\left(p,\!q,\!q''\!,\!r\right)}{\partial {\bm{{\rm \Theta}}}}\!+\!\eta_1 \!\!\!\sum_{q'\in \mathcal{N}_q} \!\!\!\frac{\partial L\!\left(p,\!q,\!q'\!,\!r\right)}{\partial {\bm{{\rm \Theta}}}} \nonumber\\
	&+\eta_2 \!\! \sum_{q'\in \mathcal{N}_q} \sum_{q''\in \mathcal{Q}_{pr}^-\setminus \mathcal{N}_q} \!\!\!\!\!\!\!\!\! \frac{\partial L\!\left(p,\!q',\!q''\!,\!r\right)}{\partial {\bm{{\rm \Theta}}}}\Bigg] \! - \! \lambda_r {\bm{{\rm \Theta}}}.
	\end{flalign}
\end{small}where
\begin{small}
	\begin{flalign}
	\frac{\partial L\!\left(p,q,q',r\right)}{\partial {\bm{{\rm \Theta}}}} = &\sigma \Big(\! -\hat{\bm{{\rm A}}}_{pqq'r} \Big)\frac{\partial \hat{\bm{{\rm A}}}_{pqq'r}}{\partial {\bm{{\rm \Theta}}}} \nonumber\\
	+ & \lambda_c \Big[ \sigma\Big(\! -\hat{\bm{{\rm B}}}_{pqq'}\Big) \frac{\partial \hat{\bm{{\rm B}}}_{pqq'}}{\partial {\bm{{\rm \Theta}}}} + \, \sigma\Big(\! -\hat{\bm{{\rm C}}}_{rqq'}\Big) \frac{\partial \hat{\bm{{\rm C}}}_{rqq'}}{\partial {\bm{{\rm \Theta}}}} \Big] \,.\nonumber
	\end{flalign}
\end{small}We use $\bm \uptheta$ to denote certain column of ${\bm{{\rm \Theta}}}$. For our VRA model, the derivatives are:
\begin{small}
	\begin{flalign}
	\label{derivatives1}
	\frac{\partial \hat{\bm{{\rm A}}}_{pqq'r}}{\partial {\bm{{\rm \uptheta}}}} = \left\{
	\begin{array}{lcl}
	{\hat{\bm{{\rm C}}}_{rq} {\bm{{\rm V}}}_{*q} - \hat{\bm{{\rm C}}}_{rq'} {\bm{{\rm V}}}_{*q'}} &\text{if} &{\rm \bm \uptheta} = {\bm{{\rm U}}}_{*p} \\
	{\hat{\bm{{\rm C}}}_{rq} {\bm{{\rm U}}}_{*p} / -\hat{\bm{{\rm C}}}_{rq'} {\bm{{\rm U}}}_{*p}} &\text{if} &{\rm \bm \uptheta} = {\bm{{\rm V}}}_{*q} / {\bm{{\rm V}}}_{*q'} \\
	{\hat{\bm{{\rm C}}}_{rq} {\bm{{\rm F}}}_{*q} - \hat{\bm{{\rm C}}}_{rq'} {\bm{{\rm F}}}_{*q'}} &\text{if} &{\rm \bm \uptheta} = {\bm{{\rm M}}}_{*p} \end{array}  
	\right.
	\end{flalign}
	
	\begin{equation}
	\label{derivatives2}
	\frac{\partial \hat{\bm{{\rm B}}}_{pqq'}}{\partial {\rm \bm \uptheta}} = \left\{
	\begin{array}{lcl}
	{{\bm{{\rm V}}}_{*q} - {\bm{{\rm V}}}_{*q'}} &\text{if} &{\rm \bm \uptheta} = {\bm{{\rm U}}}_{*p} \\
	{{\bm{{\rm U}}}_{*p} / -{\bm{{\rm U}}}_{*p}} &\text{if} &{\rm \bm \uptheta} = {\bm{{\rm V}}}_{*q} / {\bm{{\rm V}}}_{*q'} \\
	{{\bm{{\rm F}}}_{*q} - {\bm{{\rm F}}}_{*q'}} &\text{if} &{\rm \bm \uptheta} = {\bm{{\rm M}}}_{*p}
	\end{array}  
	\right.
	\end{equation}
\end{small}Equations (\ref{derivatives1}) and (\ref{derivatives2}) give the derivatives for ${\bm{{\rm \Theta}}} = \{{\bm{{\rm U}}}, {\bm{{\rm V}}}, {\bm{{\rm M}}}\}$, and we can easily get the same form for ${\bm{{\rm \Theta}}} = \{{\bm{{\rm T}}}, {\bm{{\rm W}}}, {\bm{{\rm N}}}\}$. $\frac{\partial \hat{\bm{{\rm A}}}_{pqq'r}}{\partial {\rm \bm \uptheta}}$ in Equation (\ref{derivatives1}) is certain column of $\frac{\partial \hat{\bm{{\rm A}}}_{pqq'r}}{\partial {\bm{{\rm \Theta}}}}$ in Equation (\ref{derivatives}), for example, the $p$-th column when ${\rm \bm \uptheta} = {\bm{{\rm U}}}_{*p}$.

Finally, we update the parameters with the derivatives we get. As discussed in Subsection \ref{subsec:nei_set}, we use different neighborhood sets to update different parts of the model. For ${\bm{{\rm \Theta}}} = \{{\bm{{\rm U}}}, {\bm{{\rm V}}}, {\bm{{\rm M}}}\}$, we update the parameters:
\begin{small}
	\begin{flalign}
	{\bm{{\rm \Theta}}} = {\bm{{\rm \Theta}}}+\eta \nabla_{\bm{{\rm \Theta}}} {\rm APLR\_O{\scriptstyle PT}} \big\vert_{\mathcal{N}_q = \mathcal{N}_q^U \bigcup \mathcal{N}_q^C \bigcup \mathcal{N}_q^A} \,,\nonumber
	\end{flalign}and for ${\bm{{\rm \Theta}}} = \{{\bm{{\rm T}}}, {\bm{{\rm W}}}, {\bm{{\rm N}}}\}$,
	\begin{flalign}
	{\bm{{\rm \Theta}}} = {\bm{{\rm \Theta}}}+\eta \nabla_{\bm{{\rm \Theta}}} {\rm APLR\_O{\scriptstyle PT}} \big\vert_{\mathcal{N}_q = \mathcal{N}_q^T \bigcup \mathcal{N}_q^C \bigcup \mathcal{N}_q^A} \,.\nonumber
	\end{flalign}
\end{small}Our model is optimized with mini-batch gradient descent and for each positive sample, we sample $\rho$ negative samples and $\rho$ neighbors randomly to construct pairs, where $\rho$ is the sampling rate.

\begin{algorithm}[ht]
	\caption{Learning VRA by APLR.}
	\label{alg:learning}
	\LinesNumbered
	\KwIn{sparse tensor $\bm{{\rm A}}$, coupled matrices $\bm{{\rm B}}$ and $\bm{{\rm C}}$, visual features $\bm{{\rm F}}$, weight coefficient for coupled matrices $\lambda_c$, regularization coefficient $\lambda_r$, weighting parameters $\eta_1$ and $\eta_2$, batch size $b$, learning rate $\eta$, sample rate $\rho$, maximum number of iterations $iter\_max$, and convergence criteria.}
	\KwOut{top-$n$ prediction given by the complete tensor $\hat{\bm{{\rm A}}}$.}
	construct $\mathcal{N}_q^C$, $\mathcal{N}_q^A$, $\mathcal{N}_q^U$, and $\mathcal{N}_q^T$ for each item $q$\;
	initialize $\bm{{\rm \Theta}}$ randomly\;
	$iter = 0$\;
	\While{not converged $\&\&$ $iter < iter\_max$}{
		$iter += 1$\;
		split all purchase records into $b$-size batches\;
		\For{each batch}{
			\For{each record in current batch}{
				$\mathcal{N}_q=\mathcal{N}_q^C \bigcup \mathcal{N}_q^A \bigcup \mathcal{N}_q^U$\;
				select $\rho$ neighbour items $q'$ randomly from $\mathcal{N}_q$\;
				select $\rho$ neighbour items $q''$ randomly from $\mathcal{Q}_{pr}^- \setminus \mathcal{N}_q$\;
				calculate and accumulate $\nabla_{\{{\bm{{\rm U}}},{\bm{{\rm V}}},{\bm{{\rm M}}}\}} {\rm APLR\_O{\scriptstyle PT}}$\;
				$\mathcal{N}_q=\mathcal{N}_q^C \bigcup \mathcal{N}_q^A \bigcup \mathcal{N}_q^T$\;
				select $\rho$ neighbour items $q'$ randomly from $\mathcal{N}_q$\;
				select $\rho$ neighbour items $q''$ randomly from $\mathcal{Q}_{pr}^- \setminus \mathcal{N}_q$\;
				calculate and accumulate $\nabla_{\{{\bm{{\rm T}}},{\bm{{\rm W}}},{\bm{{\rm N}}}\}} {\rm APLR\_O{\scriptstyle PT}}$\;
			}
			$\{{\bm{{\rm U}}},{\bm{{\rm V}}},{\bm{{\rm M}}}\} \,+\!\!= \eta \nabla_{\{{\bm{{\rm U}}},{\bm{{\rm V}}},{\bm{{\rm M}}}\}} \rm BPR\_O\scriptstyle PT$\;
			$\{{\bm{{\rm T}}},{\bm{{\rm W}}},{\bm{{\rm N}}}\} \,+\!\!= \eta \nabla_{\{{\bm{{\rm T}}},{\bm{{\rm W}}},{\bm{{\rm N}}}\}} \rm BPR\_O\scriptstyle PT$\; 
		}
		calculate $\hat{{\bm{{\rm A}}}}$ and predict the top-$n$ items\;
	}
	\Return{the top-$n$ items}\;
\end{algorithm}

The detailed learning procedures about our method are shown in Algorithm~\ref{alg:learning}. We first construct $\mathcal{N}_q^C$ and $\mathcal{N}_q^A$ by clustering the CNN features and aesthetic features, and construct $\mathcal{N}_q^U$ and $\mathcal{N}_q^T$ by collecting neighbours in user-item and time-item graph (line 1). We then exploit the mini-batch gradient descent to maximize the objective function. For each iteration, all positive samples are enumerated (lines 4-19). We compute the gradients with a batch containing $b$ positive samples (line 6), and select $\rho$ neighbours and $\rho$ negative samples (lines 8-16) construct preference pairs. Different parts of the model are updated with different samples (lines 12, 16, 17, and 18). To calculate the gradients (line 10), we combine Equation (\ref{derivatives}) with Equations (\ref{derivatives1}) and (\ref{derivatives2}). One thing needs to be point out is that $\frac{\partial \hat{\bm{{\rm A}}}_{pqq'r}}{\partial {\bm{{\rm \Theta}}}}$ in Equation (\ref{derivatives1}) is a certain column of $\frac{\partial \hat{\bm{{\rm A}}}_{pqq'r}}{\partial {\rm \bm \uptheta}}$ in Equation (\ref{derivatives}), for example, the $p$-th column when ${\rm \bm \uptheta} = {\bm{{\rm U}}}_{*p}$.

As we know, a more popular item tells us less about user's preference, and our optimization can weaken the contribution of popular items by nature. For a popular item $q_1$ and a minority item $q_2$, $|\mathcal{N}_{q_1}| \gg |\mathcal{N}_{q_2}|$, where $|\;\;|$ is the set size. Noting that a popular item connects to a large proportion of items, $\mathcal{N}_{q_1}$ contains various items and depicts little about the preference, and we only select a small proportion of $\mathcal{N}_{q_1}$ ($\frac{\rho}{|\mathcal{N}_{q_1}|}$). For a minority item $q_2$, $\mathcal{N}_{q_2}$ only contains similar items which the current user may prefer, therefore we select a large proportion of $\mathcal{N}_{q_2}$ ($\frac{\rho}{|\mathcal{N}_{q_2}|}$). If $q_2$ is very unpopular, we can almost cover $\mathcal{N}_{q_2}$ by sampling $\rho$ samples.

Another advantage of our neighbour-enhanced pairwise optimization is that important neighbours can be strengthened in the probability level. We give an example to illustrate this advantage.

\begin{figure}[ht!]
	\centering
	\includegraphics[scale = 0.6]{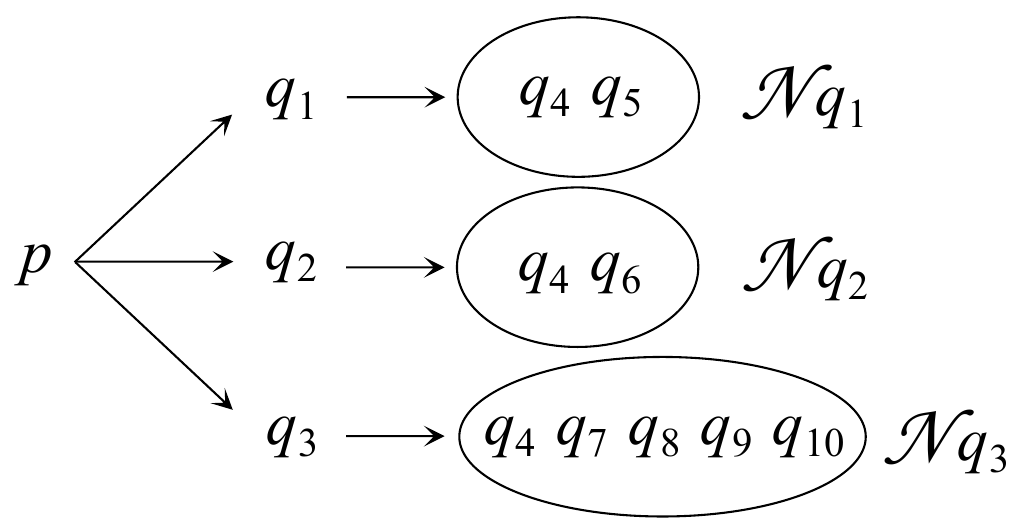}
	\caption{An example of neighbour sets.}
	\label{fig:item_collaborative}
\end{figure}
\begin{myEXP}
	
	As shown in Figure~\ref{fig:item_collaborative}, $p$ is the current user, and $q_1$, $q_2$, $q_3$ are three positive samples. $q_{4-10}$ are neighbours of positive samples and with high probability to be preferred by $p$, hence are potential positive samples. As we can see, $q_4$ is the most important potential items since it is the neighbour of all $p$'s purchased items and with the highest probability to be preferred. $q_{7-10}$ are not important since they are the neighbours of $q_3$, which is a popular item. As we discussed, $\mathcal{N}_{q_3}$ provides little information about $p$'s preference. When sampling from the neighbour set, taking $\rho = 1$ as an example, $q_4$ has $\frac{1}{20}$, $\frac{3}{10}$, $\frac{9}{20}$, and $\frac{1}{5}$ probability to be sampled 3 times, twice, once, and not to be sampled, respectively. $q_5$ and $q_6$ both have $\frac{1}{2}$ probability to be sampled (once) and have $\frac{1}{2}$ probability not to be sampled. $q_{7-10}$ all have $\frac{1}{5}$ probability to be sampled (once) and have $\frac{4}{5}$ probability not to be sampled. Assuming we iterate 200 times to train our model, $q_4$ can be sampled about 240 times, $q_5$ and $q_6$ can both be sampled about 100 times, and $q_{7-10}$ can all be sampled about 40 times. We can see that potential samples are weighted based on the importance in the probability (frequency) level. To improve the sampling quality, \citet{CPLR} weighted potential samples based on the strength of the connection yet additional computation is required. Compared with \cite{CPLR}, our method weights potential samples by nature.
\end{myEXP}

\section{Experiments}
\label{sec:experiments}
In this section, we conduct experiments on real-world datasets to verify the effectiveness of our method. We focus on answering the following four key research questions:

\vspace{2mm}
\noindent \textbf{RQ1:} What factors affect users' aesthetics?
\vspace{1mm}

\noindent \textbf{RQ2:} How is the performance of our overall solution for the clothing recommendation task?
\vspace{1mm}

\noindent \textbf{RQ3:} How is the effectiveness of the aesthetic features compared with conventional visual features?
\vspace{1mm}

\noindent \textbf{RQ4:} How is the performance of our aesthetic-enhanced learning to rank method?
\vspace{1mm}

\subsection{Experimental Setup}
\subsubsection{Datasets}
We use the \emph{AVA} dataset to train the aesthetic network and use the \emph{Amazon} dataset to train the recommendation models.
\begin{itemize}
	\item{\textbf{Aesthetic Visual Analysis (AVA):} We train the aesthetic network with the \emph{AVA} dataset \cite{AVA}, which is the collection of images and meta-data derived from \emph{DPChallenge.com}. It contains over 250,000 images with aesthetic ratings from 1 to 10, 66 textual tags describing the semantics of images, and 14 photographic styles: complementary colors, duotones, negative image, rule of thirds, image grain, silhouettes, vanishing point, high dynamic range, light on white, long exposure, macro, motion blur, shallow DOF, and soft focus. We abandon the last 7 styles when constructing pathways in our aesthetic feature extractor since they are about camera setting.}
	
	\item{\textbf{Amazon:} The \emph{Amazon} dataset \cite{VBPR} is the consumption records from \emph{Amazon.com}. In this paper, we use the \emph{clothing shoes and jewelry} category filtered with \emph{5-core} (remove users and items with less than 5 purchase records) to train all recommendation models. Please note that in the below part of this paper, we use \emph{Amazon} to denote the \emph{clothing shoes and jewelry} category.}
\end{itemize}

\subsubsection{Experiment Settings}
In the \emph{Amazon} dataset, we remove the record before 2010. Time is discretized by weeks, and there are 237 time intervals in total. To validate the scalability of the model and give a comprehensive assessment, we split the dataset into several subsets by gender and categories of products (\textit{Jewelry} dataset includes both jewelries and watches).

\begin{table}[ht!]
	\caption{Statistics of datasets.}
	\centering
	\label{tab:datasets}
	\scalebox{0.8}{
		\begin{tabular}{ccccc}
			\toprule
			Dataset & Purchase & User & Item & Sparsity of Matrices/Tensors\\
			\midrule
			\textit{Amazon} & 275539 & 39371 & 23022 & 99.9696\% / 99.9999\% \\
			\textit{Men} & 67156 & 22547 & 5460 & 99.9454\% / 99.9998\% \\
			\textit{Women} & 176136 & 35059 & 14500 & 99.9653\% / 99.9999\% \\
			\textit{Clothes} & 115841 & 32728 & 8777 & 99.9597\% / 99.9998\% \\
			\textit{Shoes} & 94560 & 32538 & 8231 & 99.9647\% / 99.9999\% \\
			\textit{Jewelry} & 37314 & 15924 & 3607 & 99.9350\% / 99.9997\% \\
			\bottomrule
	\end{tabular}}
\end{table}

We then randomly split each dataset into training (80\%), validation (10\%), and test (10\%) sets, and remove the cold items and users (items and users without records in training set) from the validation and test sets. The validation set is used for tuning hyper-parameters and the final performance comparison is conducted on the test set. The $F_1$-score and the normalized discounted cumulative gain (NDCG) are used to evaluate the performance of the baselines and our model. We recommend the top-$n$ items to each user to calculate $F_1$-score and NDCG for this user, and calculate the average score as the model performance. Our experiments are conducted by predicting Top-5, 10, 20, 50, and 100 favourite clothing.

\subsection{Influential Factors of Aesthetics (RQ1)}
In this subsection, we explore what factors impact the users' aesthetics by reporting some statistics of the low-level aesthetics features: Hue, Saturation, and Value (HSV). Here we use HSV rather than the high-level aesthetics features we extract because that the high-level features (high-dimensional vectors) are difficult to count and to represent, moreover, the specific meaning of each dimension is not clear. HSV is low-level yet representative, and makes the experiment result explainable.

\begin{figure*}[ht!]
	\centering
	\subfigure[Hue]{
		\includegraphics[scale = 0.45]{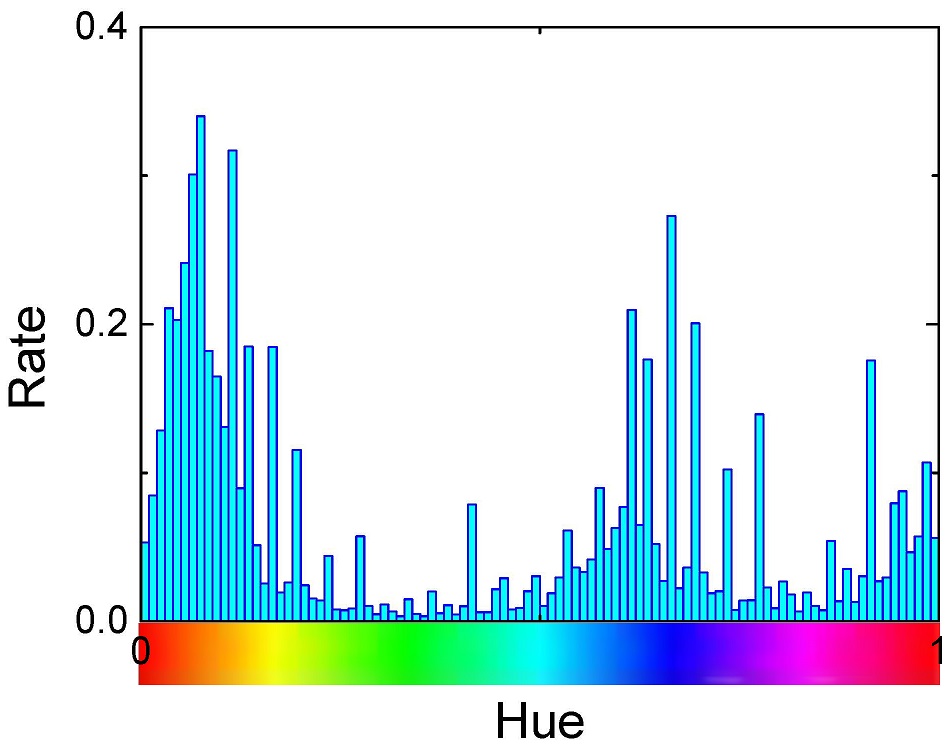}
		\label{fig:h_all}
	}
	\hspace{2mm}
	\subfigure[Saturation]{
		\includegraphics[scale = 0.45]{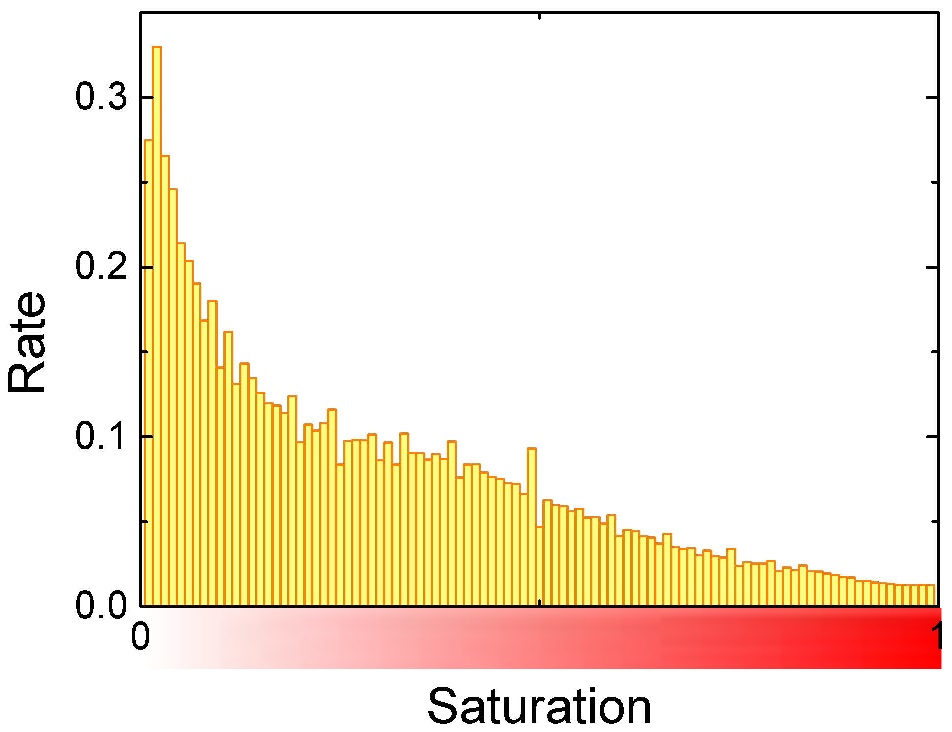}
		\label{fig:s_all}
	}
	\hspace{2mm}
	\subfigure[Value]{
		\includegraphics[scale = 0.45]{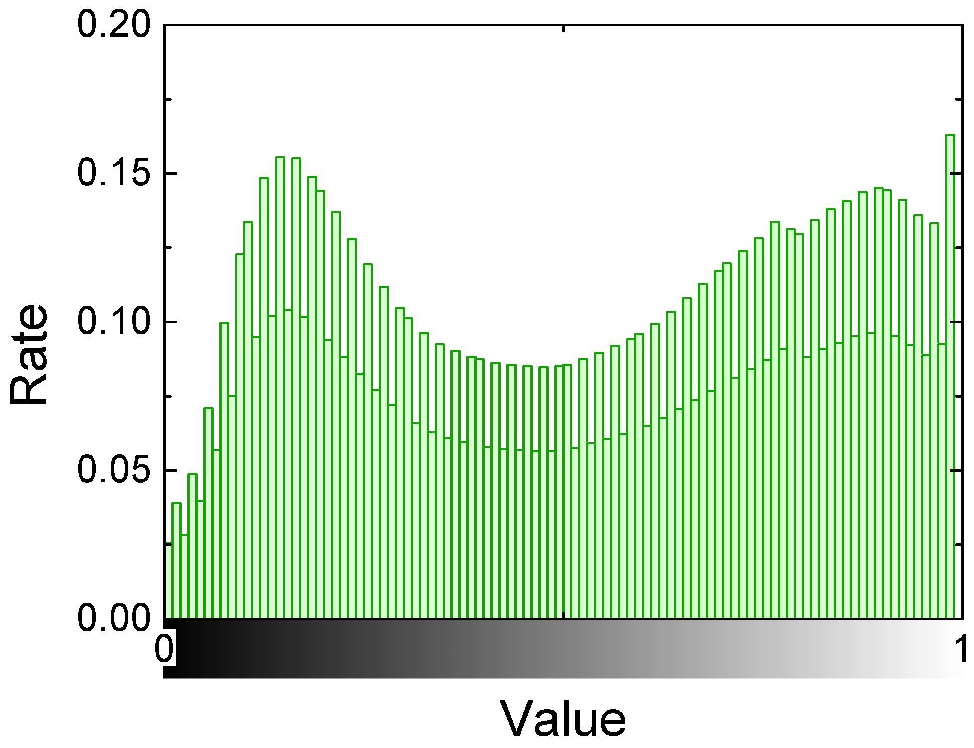}
		\label{fig:v_all}
	}
	\caption{Distribution of hue, saturation, and value of the whole dataset.}
	\label{fig:hsv_all}
\end{figure*}

Figure~\ref{fig:hsv_all} shows the distribution of hue, saturation, and value, which are counted from the whole \textit{Amazon} dataset (the \emph{clothing shoes and jewelry} category). We normalize hue, saturation, and value into $[0,1]$ and normalize the histograms into a unit vector. The bar in the bottom of Figure~\ref{fig:h_all} is the hue, and different hue indicates different colors. From the figure we can see that users prefer red and blue. The bar in the bottom of Figure~\ref{fig:s_all} is the saturation, which defines the brilliance and intensity of a color. From Figure~\ref{fig:s_all}, we can see that users prefer a lower saturation. The bar in the bottom of Figure~\ref{fig:v_all} is the value, which refers to the lightness or darkness of a color. The larger the value is, the lighter the color is. To present the difference of aesthetic preferences with certain factor, we report the difference between the normalized HSV histograms before and after the influence of certain factor, so there are positive values and negative values (see Figures~\ref{fig:s_age} to~\ref{fig:h_year}). We mainly discuss the variation of HSV with different kinds of users and in different time.

\subsubsection{Influence of Users}
Modern recommender systems aim to provide the personalized recommendation, so the influence of different kinds of users is very important. It is obvious that different users have different aesthetic preferences. In this subsection, we show the variation of HSV impacted with the gender and age.

\begin{figure}[ht!]
	\centering
	\subfigure[Kids]{
		\includegraphics[scale = 0.4]{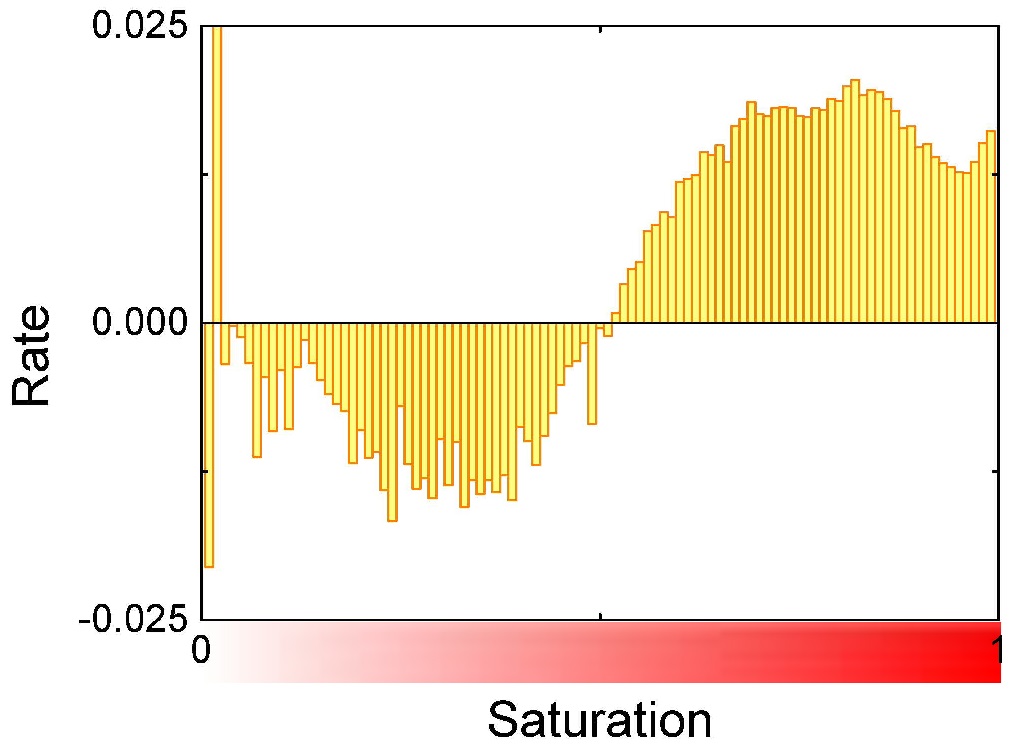}
		\label{fig:s_kid}
	}
	\hspace{-4mm}
	\subfigure[Adults]{
		\includegraphics[scale = 0.4]{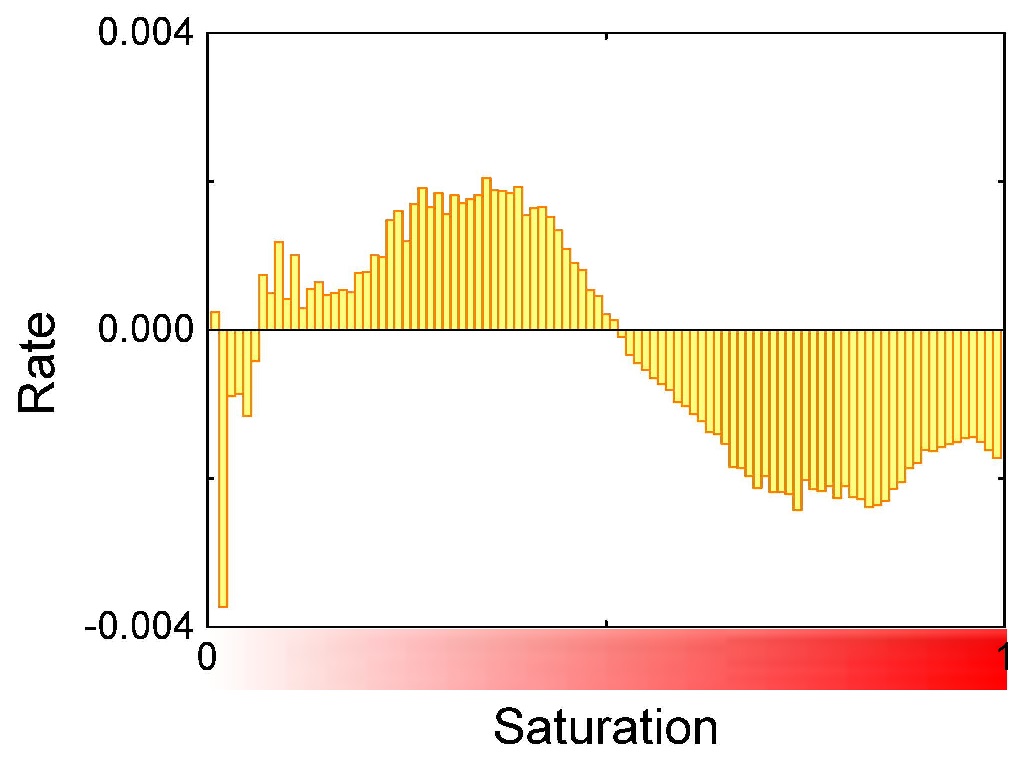}
		\label{fig:s_adults}
	}
	\caption{Aesthetic preferences of users with different ages.}
	\label{fig:s_age}
\end{figure}

\textbf{Users with different ages:} Figure~\ref{fig:s_age} shows the impact of users with different ages. Figure~\ref{fig:s_kid} and~\ref{fig:s_adults} show the saturation distribution of kids and adults, respectively. Kids like clothes with really high saturation while adults like those with low saturation.

\begin{figure}[ht!]
	\centering
	\subfigure[Men]{
		\includegraphics[scale = 0.4]{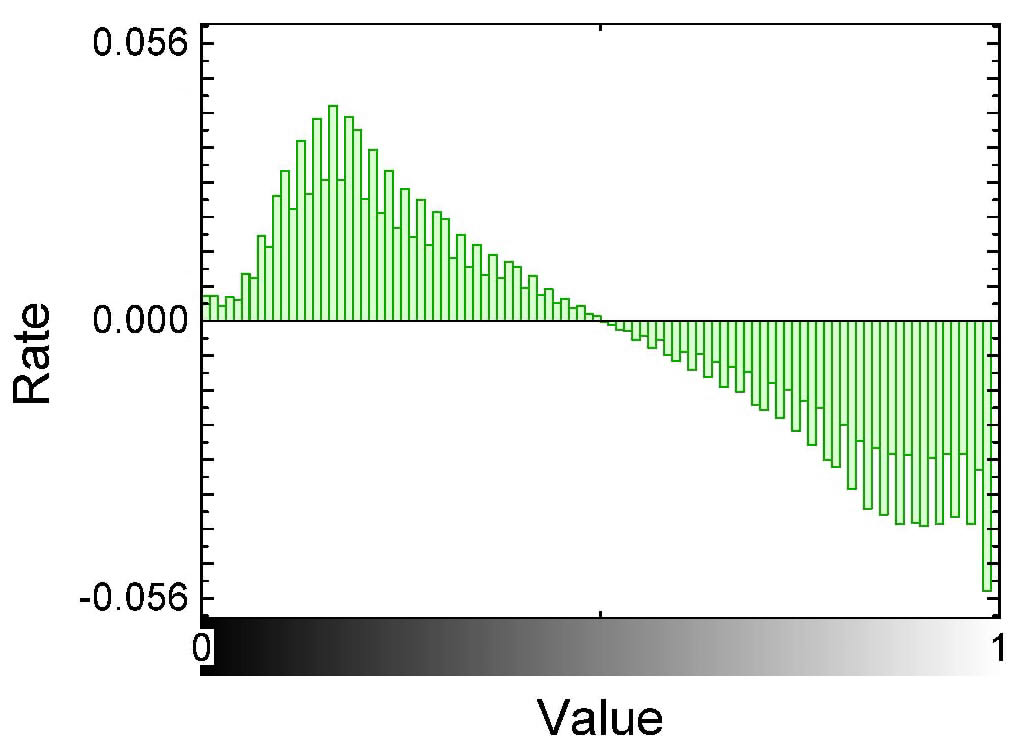}
		\label{fig:v_men}
	}
	\hspace{-4mm}
	\subfigure[Women]{
		\includegraphics[scale = 0.4]{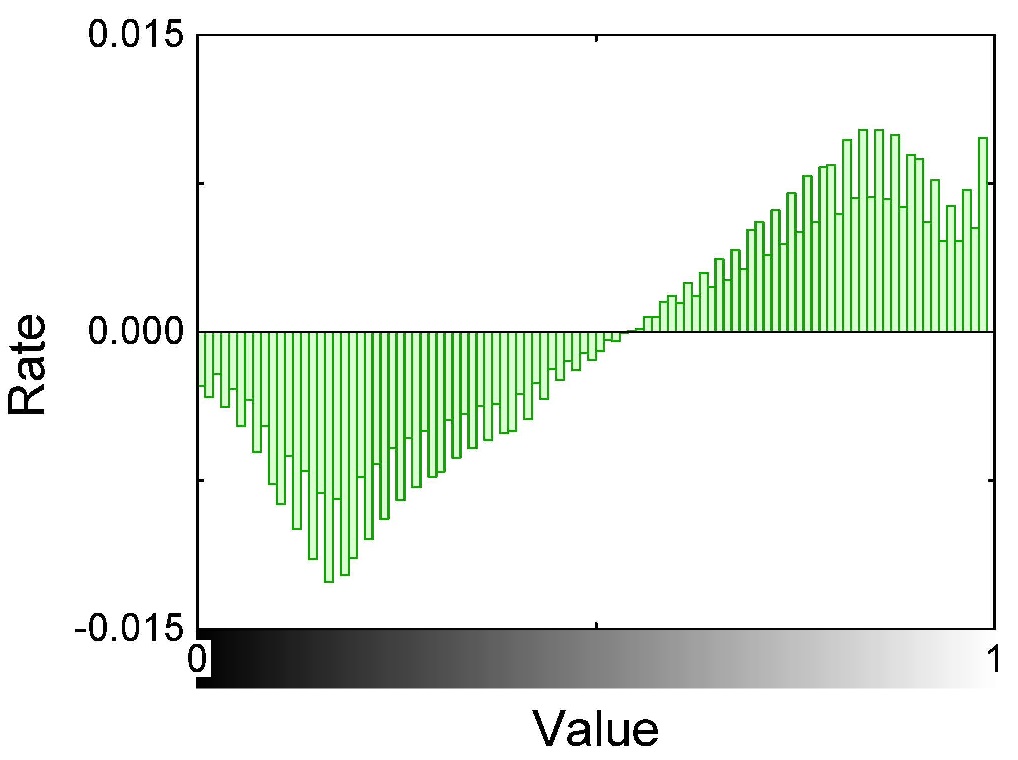}
		\label{fig:v_women}
	}
	\caption{Aesthetic preferences of users with different genders.}
	\label{fig:v_gender}
\end{figure}

\textbf{Users with different genders:} Figure~\ref{fig:v_gender} presents the aesthetic preferences of males and females. Figure~\ref{fig:v_men} shows the distribution of the value with males. They prefer dark clothes that can make them look mature and steady. Figure~\ref{fig:v_women} shows the distribution with females. They prefer lovely and active clothes in light colors. 

\subsubsection{Influence of Time}
For many products, especially clothes, movies, electronic devices, etc., sales change dramatically with time. Users' aesthetic preferences also change with time. For example, people like different colors and design in different seasons. Also, the fashion changes every year. In this subsection, we represent how time influences aesthetic preferences in a short term and long term.

\begin{figure}[ht!]
	\centering
	\subfigure[Spring]{
		\includegraphics[scale = 0.4]{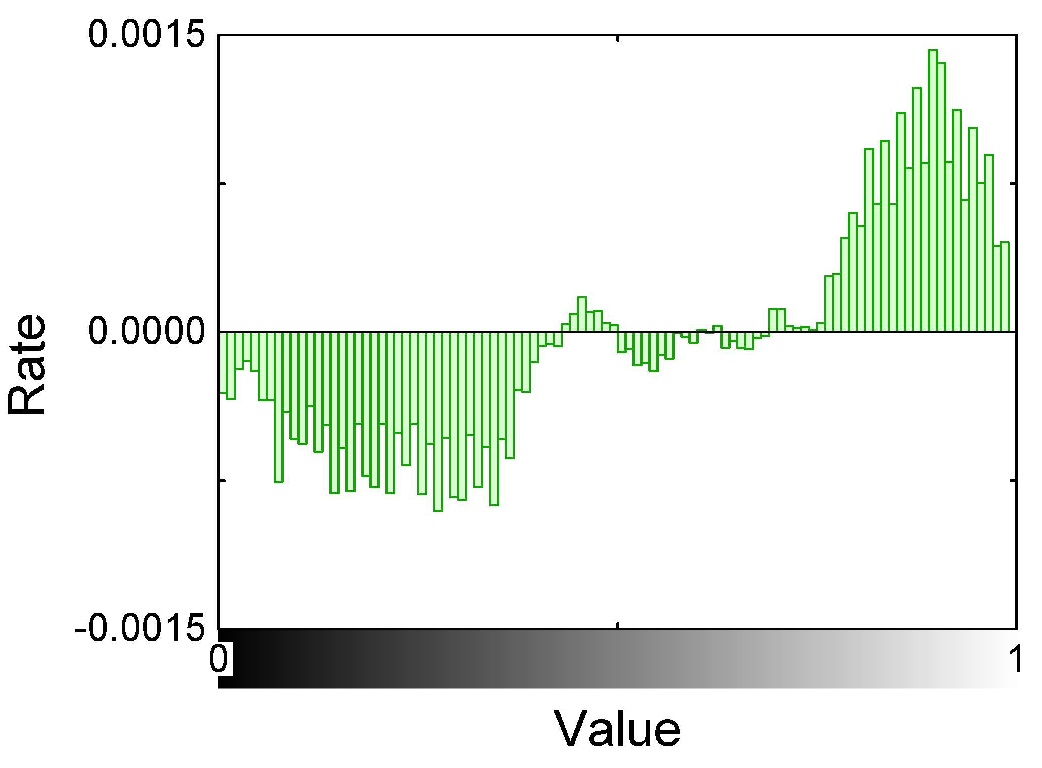}
		\label{fig:v_spring}
	}
	\hspace{-4mm}
	\subfigure[Summer]{
		\includegraphics[scale = 0.4]{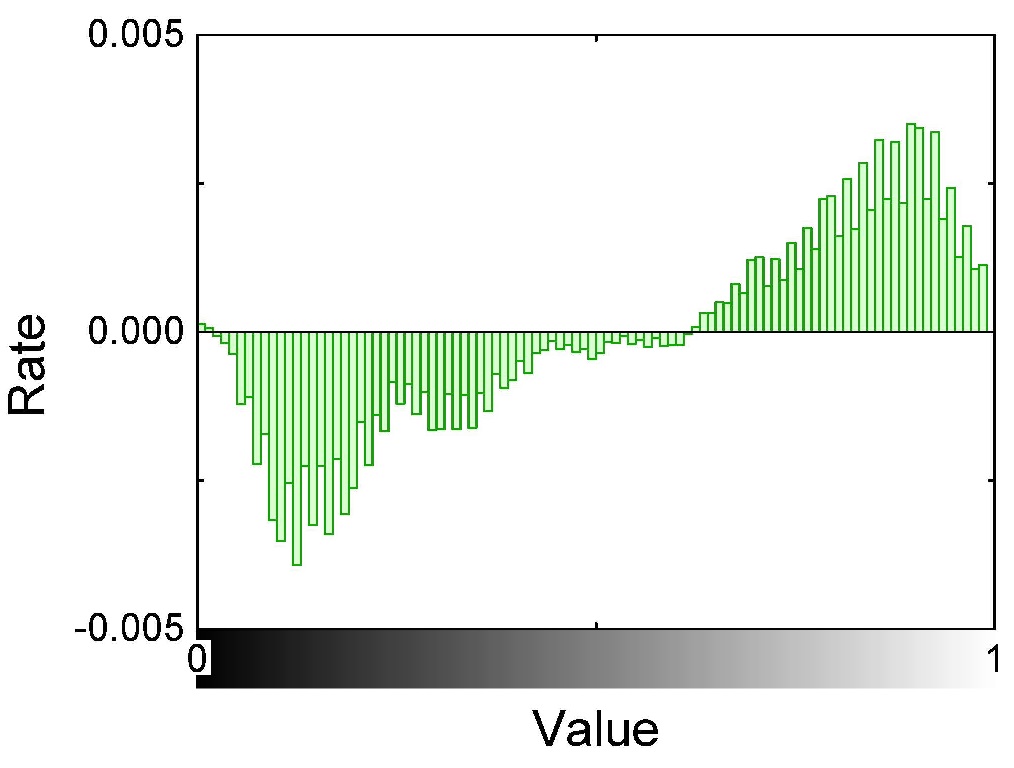}
		\label{fig:v_summer}
	}
	\hspace{-4mm}
	\subfigure[Autumn]{
		\includegraphics[scale = 0.4]{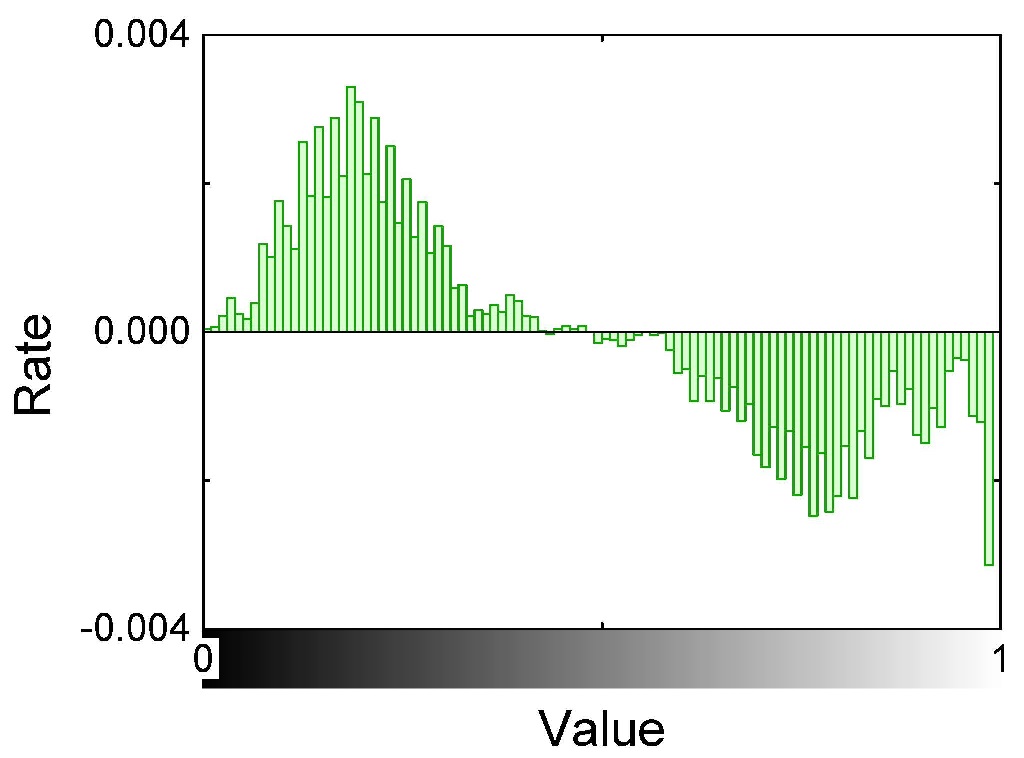}
		\label{fig:v_autumn}
	}
	\hspace{-4mm}
	\subfigure[Winter]{
		\includegraphics[scale = 0.4]{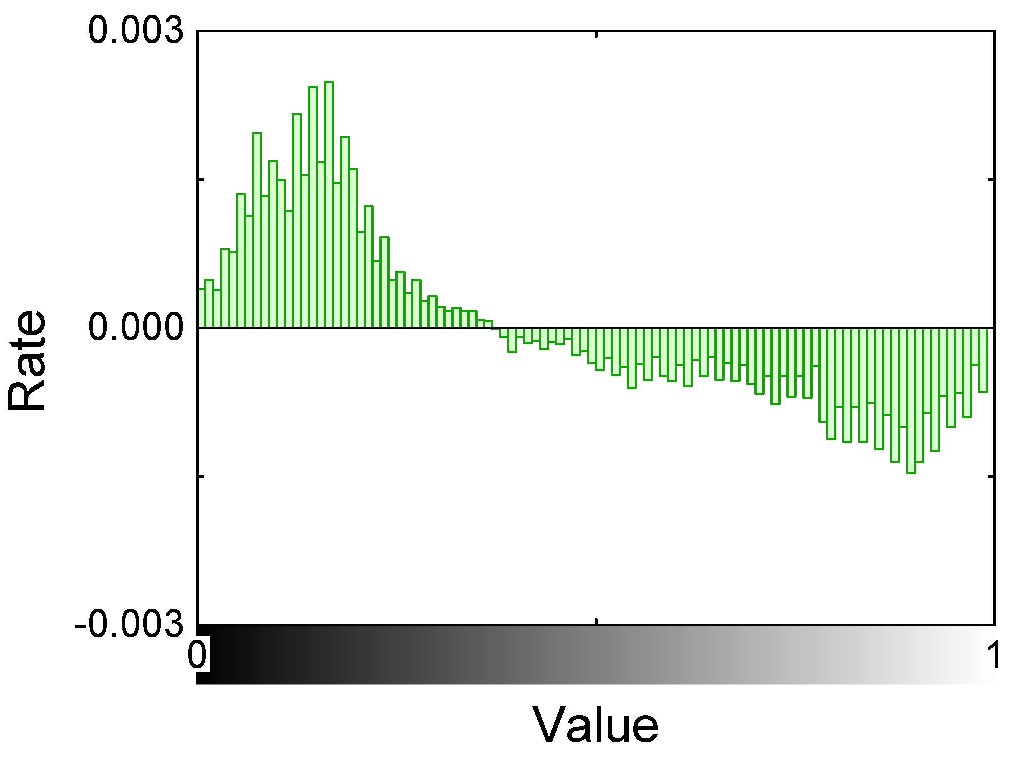}
		\label{fig:v_winter}
	}
	\hspace{-4mm}
	\caption{Aesthetic preferences in different seasons.}
	\label{fig:v_season}
\end{figure}

\textbf{Seasonality:} Figure~\ref{fig:v_season} represents users' aesthetic preferences in different seasons. Figures~\ref{fig:v_spring} to~\ref{fig:v_winter} show the distribution of value in spring, summer, autumn and winter, respectively. Users prefer light colors in spring and summer while prefer dark colors in autumn and winter.

\begin{figure*}[ht!]
	\centering
	\subfigure[2010]{
		\includegraphics[scale = 0.45]{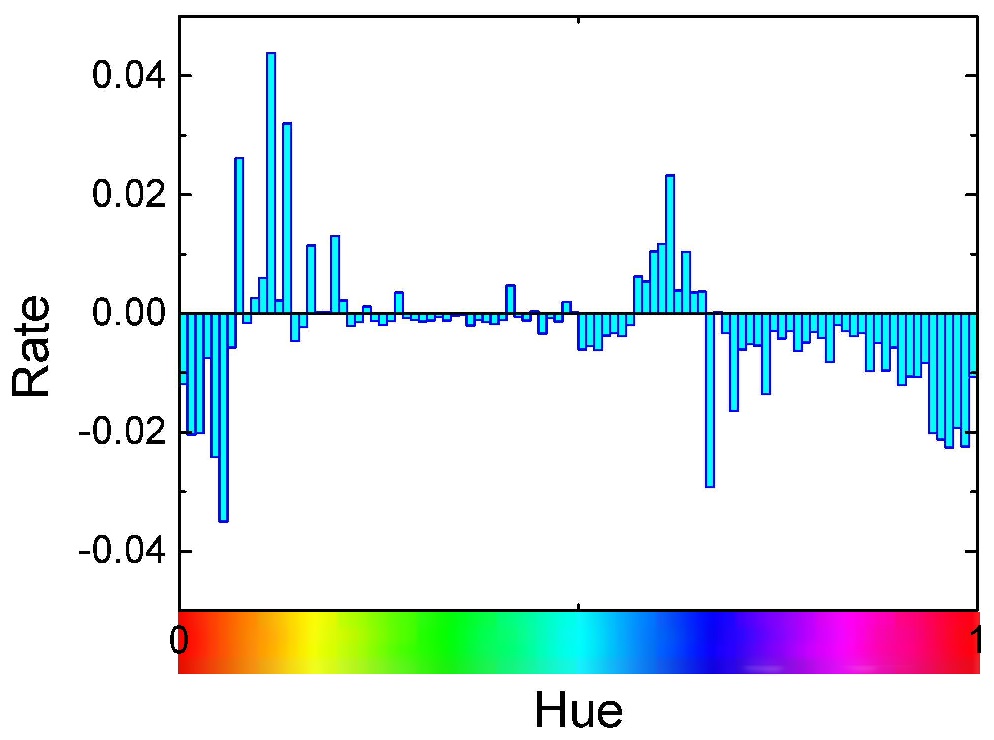}
		\label{fig:h_2010}
	}
	\hspace{2mm}
	\subfigure[2012]{
		\includegraphics[scale = 0.45]{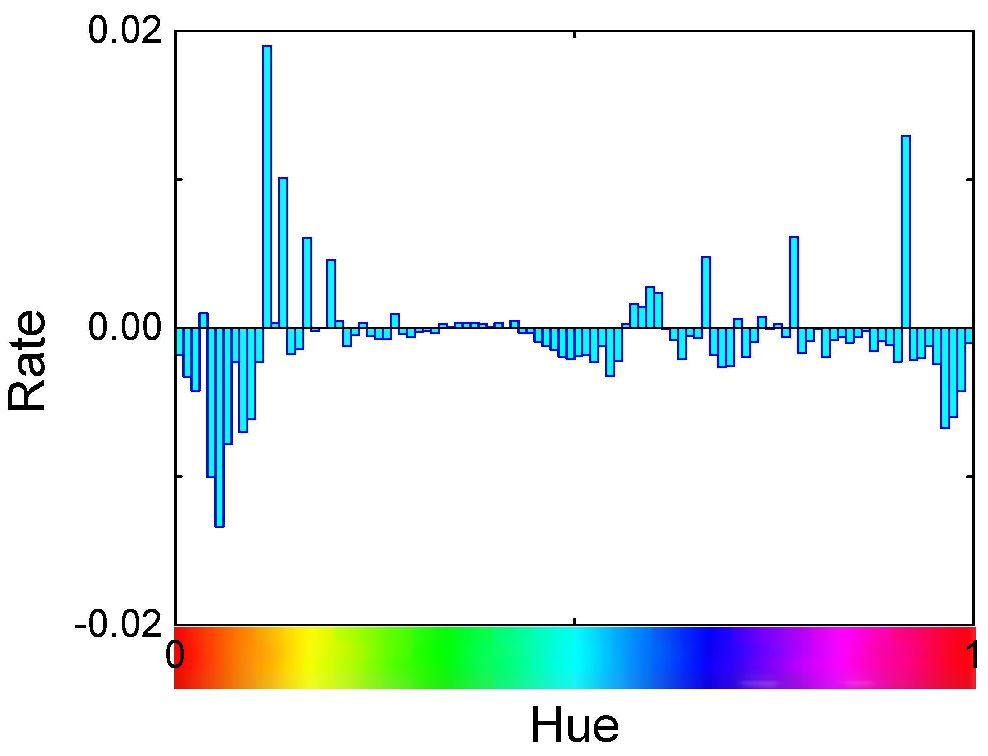}
		\label{fig:h_2012}
	}
	\hspace{2mm}
	\subfigure[2014]{
		\includegraphics[scale = 0.45]{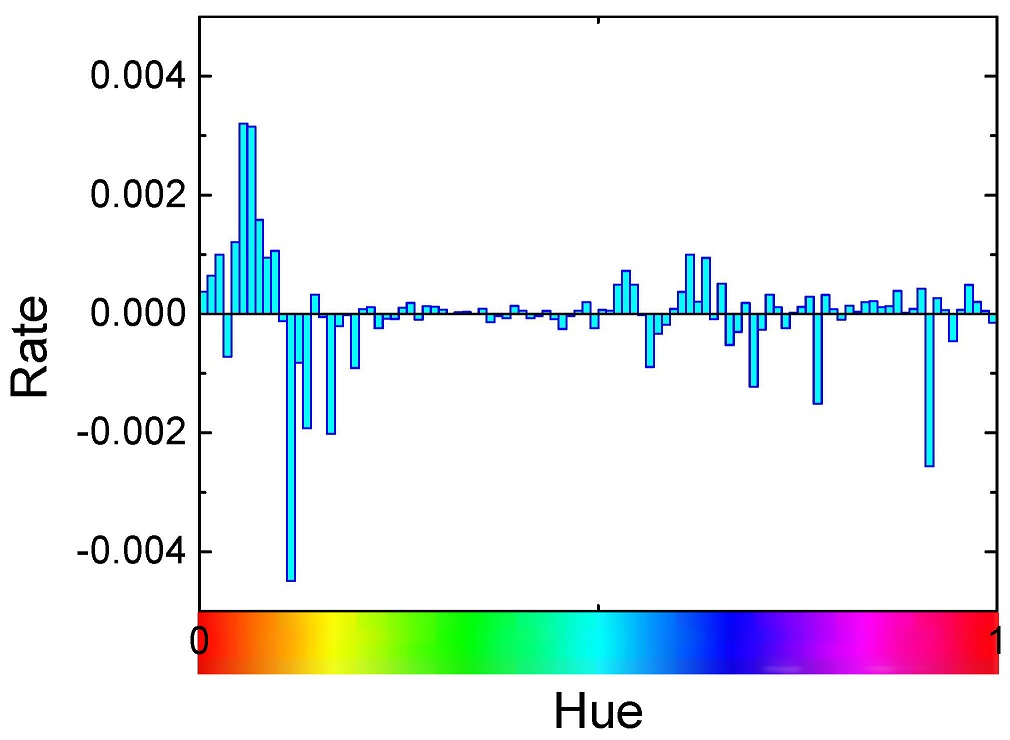}
		\label{fig:h_2014}
	}
	\caption{Aesthetic preferences in different years.}
	\label{fig:h_year}
\end{figure*}

\textbf{Annual trend:} The fashion trend in different years is shown in Figure~\ref{fig:h_year}. Histograms in Figures~\ref{fig:h_2010} to~\ref{fig:h_2014} show the hue distribution of clothes in 2010, 2012 and 2014, respectively. As shown in Figure~\ref{fig:h_year}, users preferred yellow and blue in 2010. In 2012, yellow and purple became popular. In 2014, the most popular color was red.

From the figures above, we come to the conclusion that users' aesthetic preferences change with different people and different time. So we propose a time-aware model taking these two factors into account as the basic model.

\subsection{Performance of Our Model (RQ2)}
To demonstrate the effectiveness of our model, we adopt the following methods as baselines for performance comparison:
\begin{itemize}
	
	\item{\textbf{BPR:} This \textbf{B}ayesian \textbf{P}ersonalized \textbf{R}anking method is a well known ranking-based method \cite{BPR} for implicit feedback. The preference pairs are constructed between the positive samples and the other ones. In our experiments, we optimize matrix factorization (MF) model with the pairwise optimization.}
	
	\item{\textbf{VBPR:} This \textbf{V}isual \textbf{B}ayesian \textbf{P}ersonalized \textbf{R}anking method is a visually aware recommendation method \cite{VBPR}. The visual features are pre-generated from the product images using CNN.}
	
	\item{\textbf{VNPR:} This \textbf{V}isual \textbf{N}eural \textbf{P}ersonalized \textbf{R}anking method is a visually aware neural network for recommendation  \cite{Neural_Personalize}. We predict the user preference with both embeddings and visual features. Interactions of users and items are achieved by a deep structure.}
	
	\item{\textbf{DVBPR:} This \textbf{D}eep \textbf{V}isual \textbf{B}ayesian \textbf{P}ersonalized \textbf{R}anking method is an end-to-end visually aware neural recommendation model \cite{Visually_Aware}. In \cite{Visually_Aware}, the embedding layer is removed and CNN is trained from the scratch to predict the user preference. Our experiments show that this setting is suboptimal, thus we reserve the embedding layer and pretrain the CNN on \textit{ImageNet}.}
	
	\item{\textbf{CPLR:} This \textbf{C}ollaborative \textbf{P}airwise \textbf{L}earning to \textbf{R}ank method \cite{CPLR} is an extension of BPR. Collaborative information is used to improve the quality of negative sampling and further improve the quality of ranking.}
	
	\item{\textbf{WBPR:} \textbf{W}eighted \textbf{B}ayesian \textbf{P}ersonalized \textbf{R}anking \citep{WBPR} is an extension of BPR. WBPR improves the quality of negative sampling depending on the item popularity. Considering that popular items are unlikely to be neglected, WBPR gives larger confidence weights to negative samples with higher popularity.}
		
\end{itemize}

\begin{figure*}[ht!]
	\centering
	\subfigure[\textit{Amazon}]{
		\includegraphics[scale = 0.25]{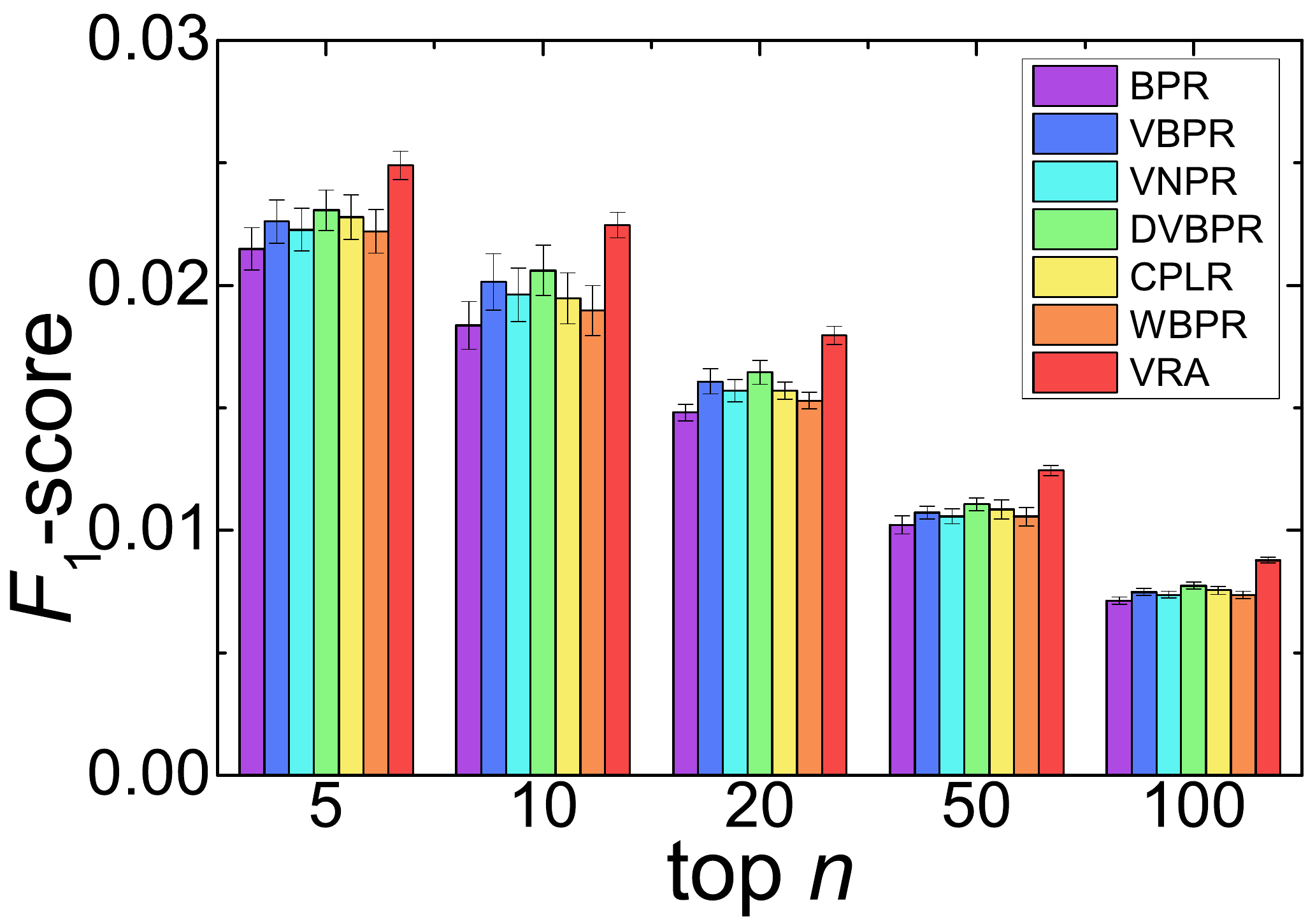}
		\label{fig:all_F1}
	}
	\hspace{-2mm}
	\subfigure[\textit{Women}]{
		\includegraphics[scale = 0.25]{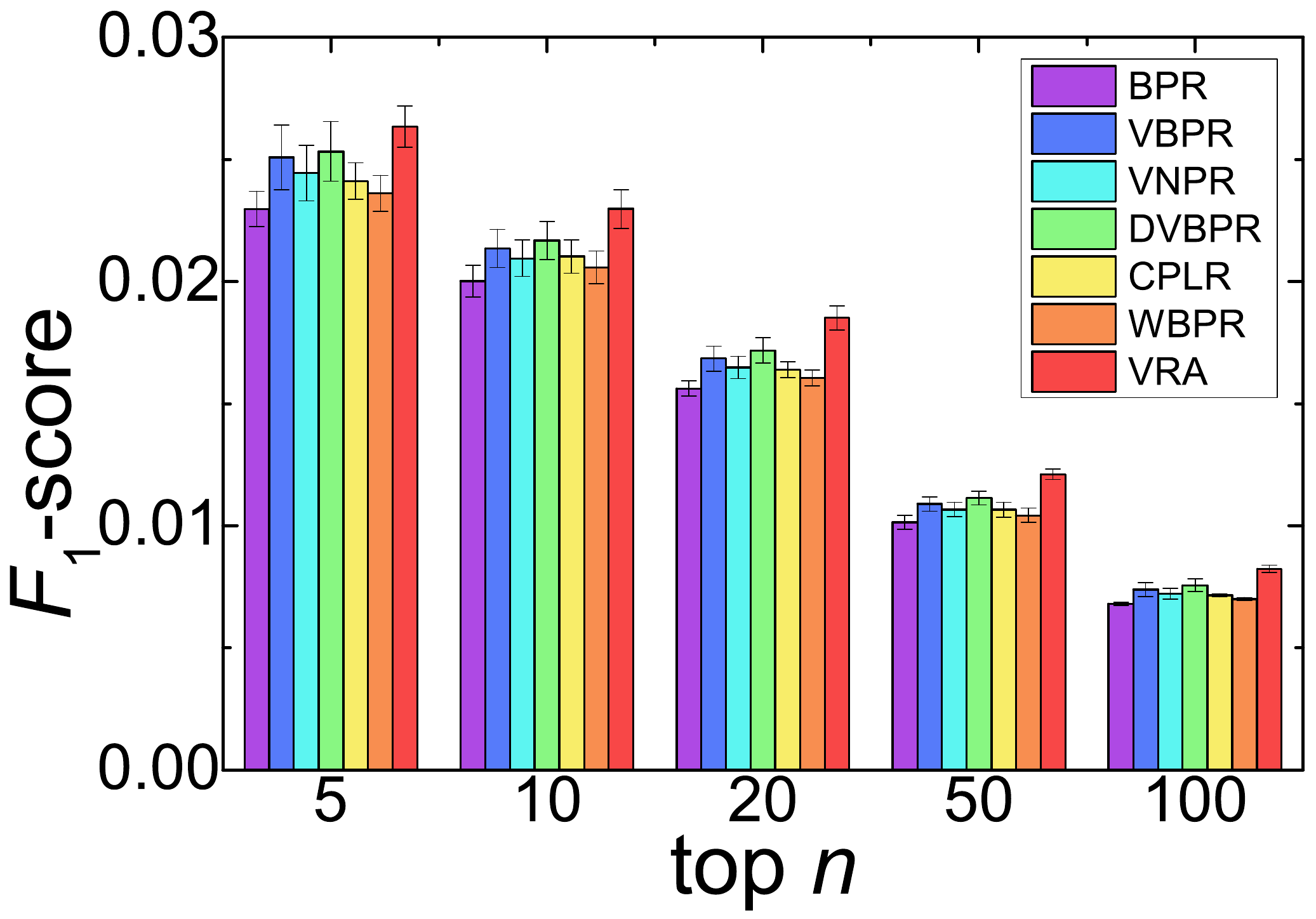}
		\label{fig:women_F1}
	}
	\subfigure[\textit{Men}]{
		\includegraphics[scale = 0.25]{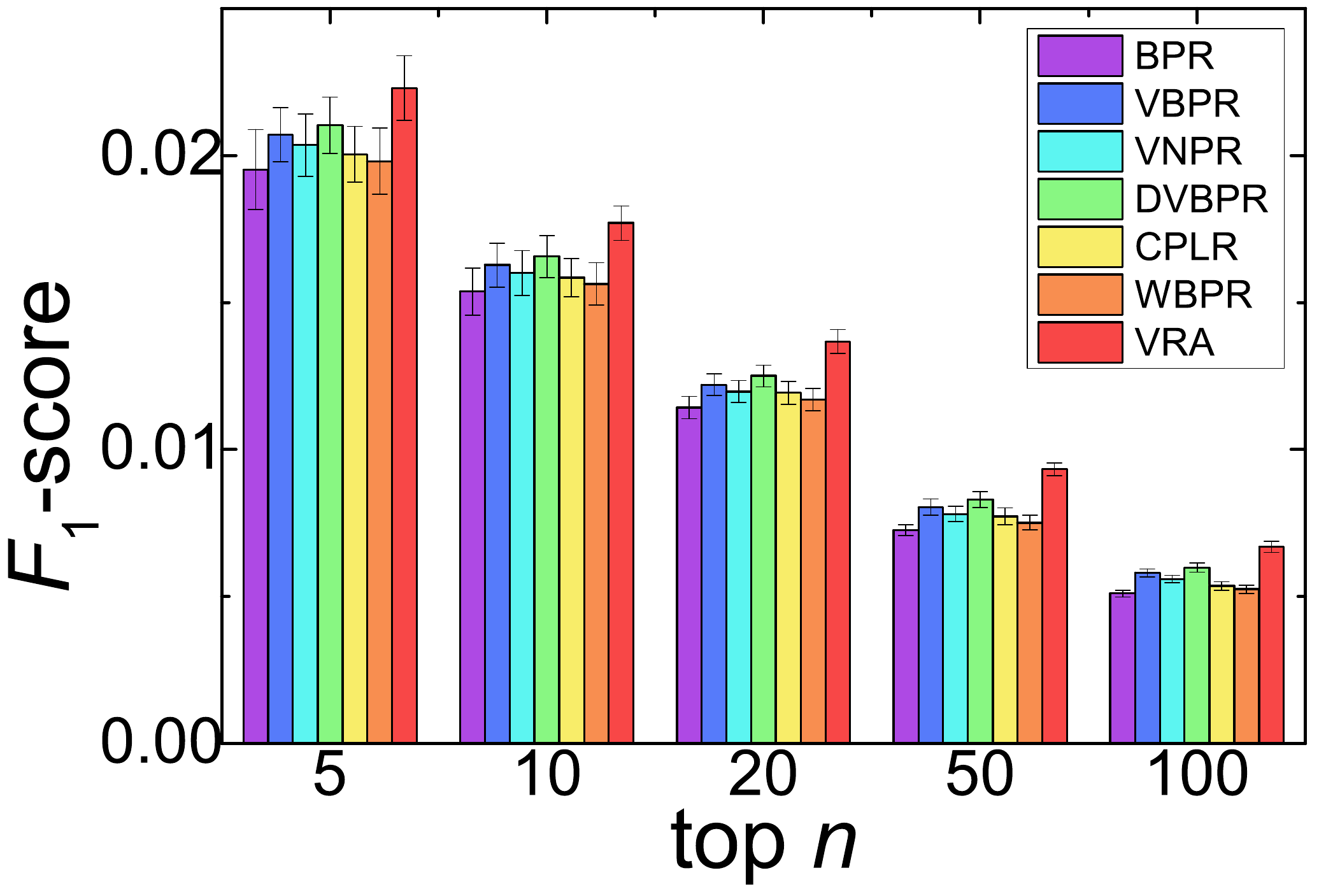}
		\label{fig:men_F1}
	}
	\hspace{-2mm}
	\subfigure[\textit{Clothes}]{
		\includegraphics[scale = 0.25]{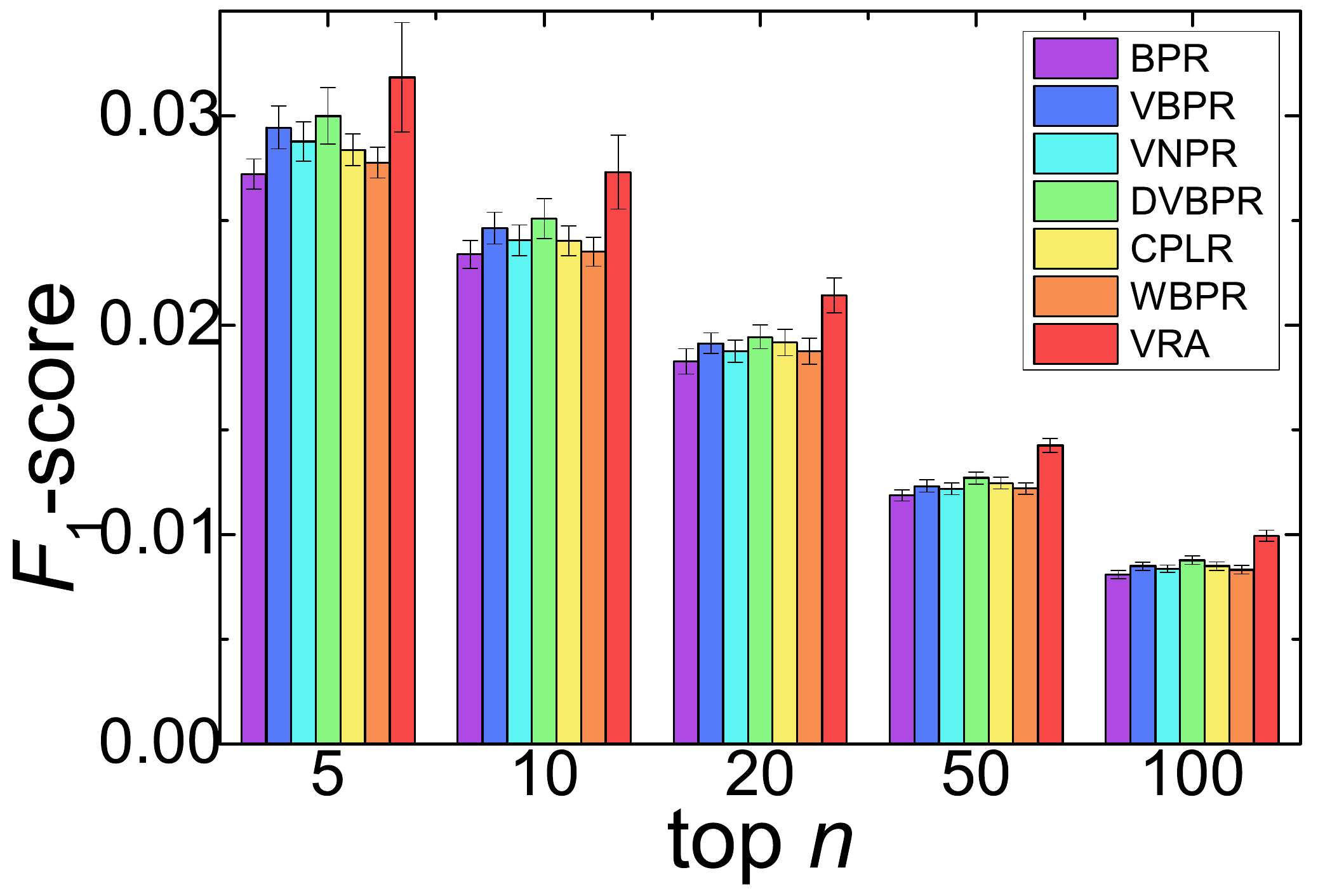}
		\label{fig:clothes_F1}
	}
	\subfigure[\textit{Shoes}]{
		\includegraphics[scale = 0.25]{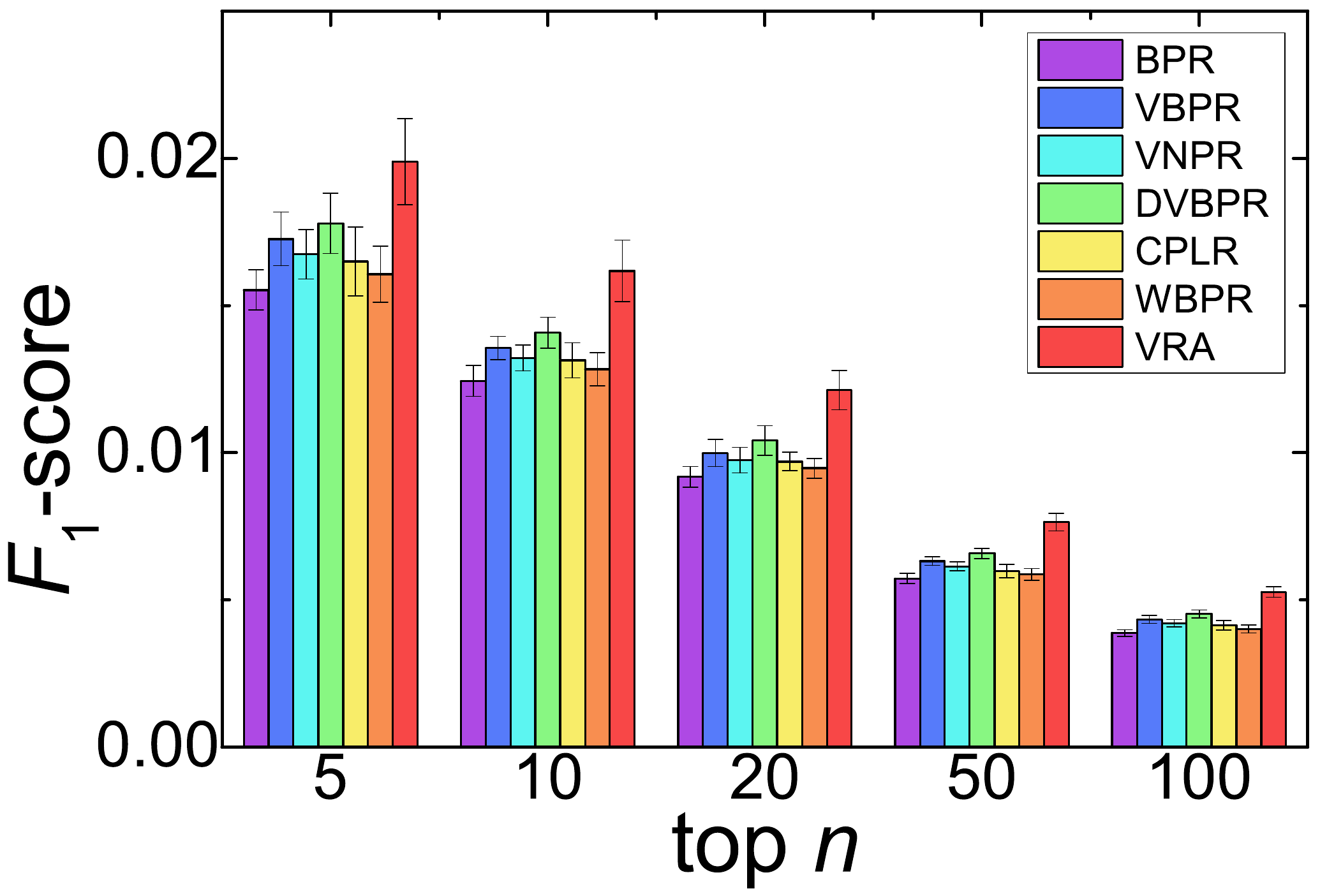}
		\label{fig:shoes_F1}
	}
	\hspace{-2mm}
	\subfigure[\textit{Jewelry}]{
		\includegraphics[scale = 0.25]{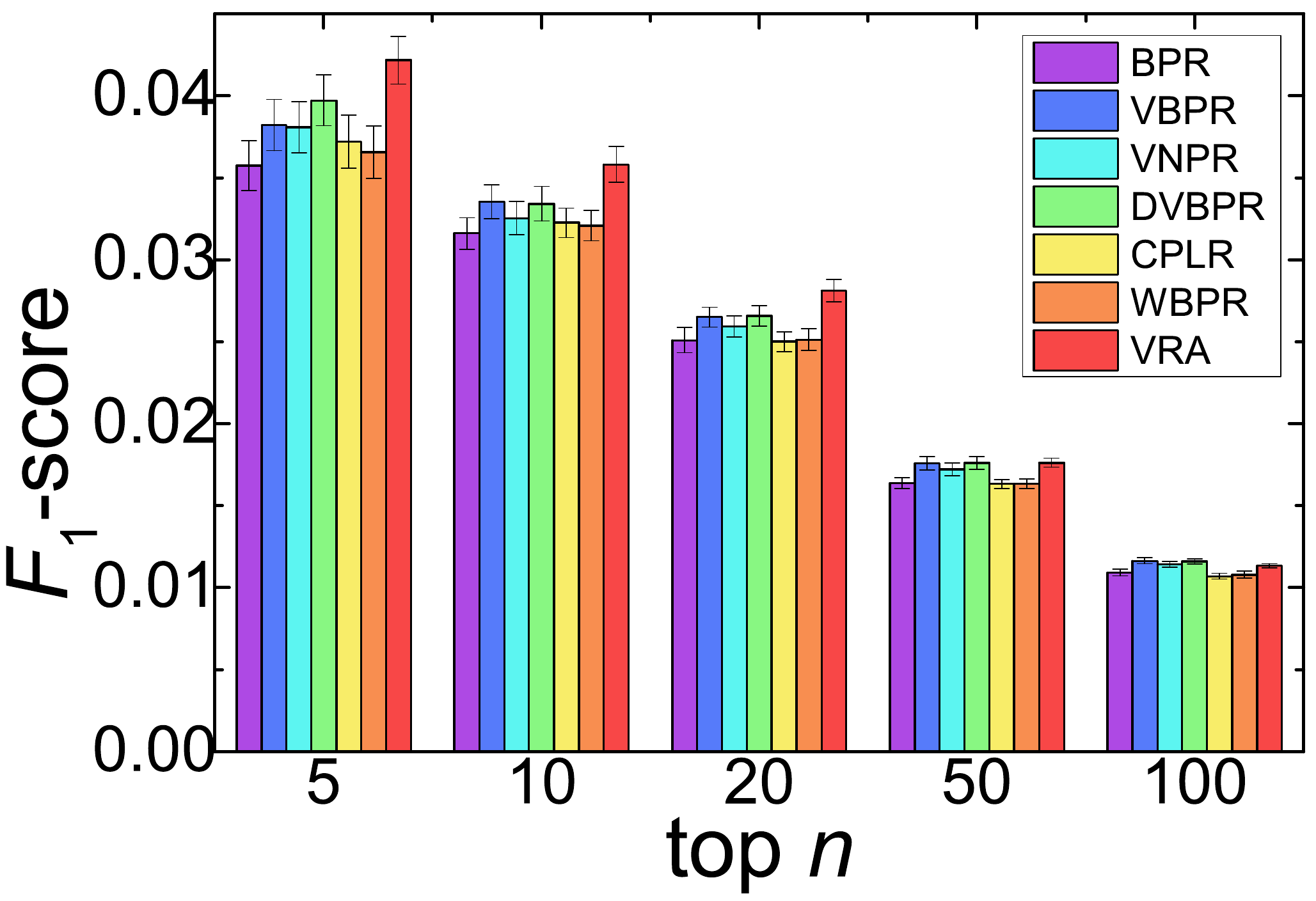}
		\label{fig:jewelry_F1}
	}
	\caption{$F_1$-score of different datasets (test set)}
	\label{fig:perf_F1}
\end{figure*}

\begin{figure*}[ht!]
	\centering
	\subfigure[\textit{Amazon}]{
		\includegraphics[scale = 0.25]{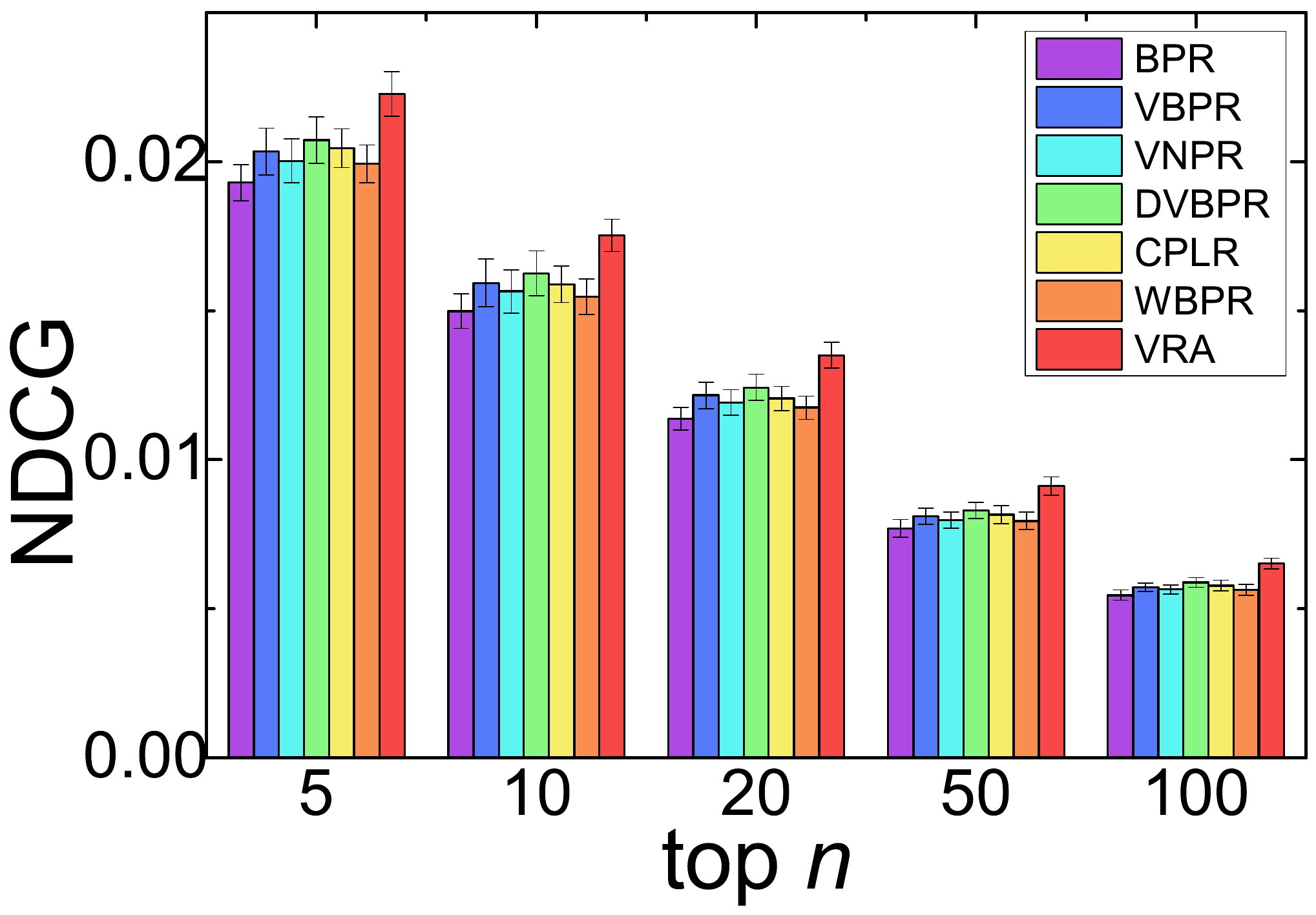}
		\label{fig:all_NDCG}
	}
	\subfigure[\textit{Women}]{
		\includegraphics[scale = 0.25]{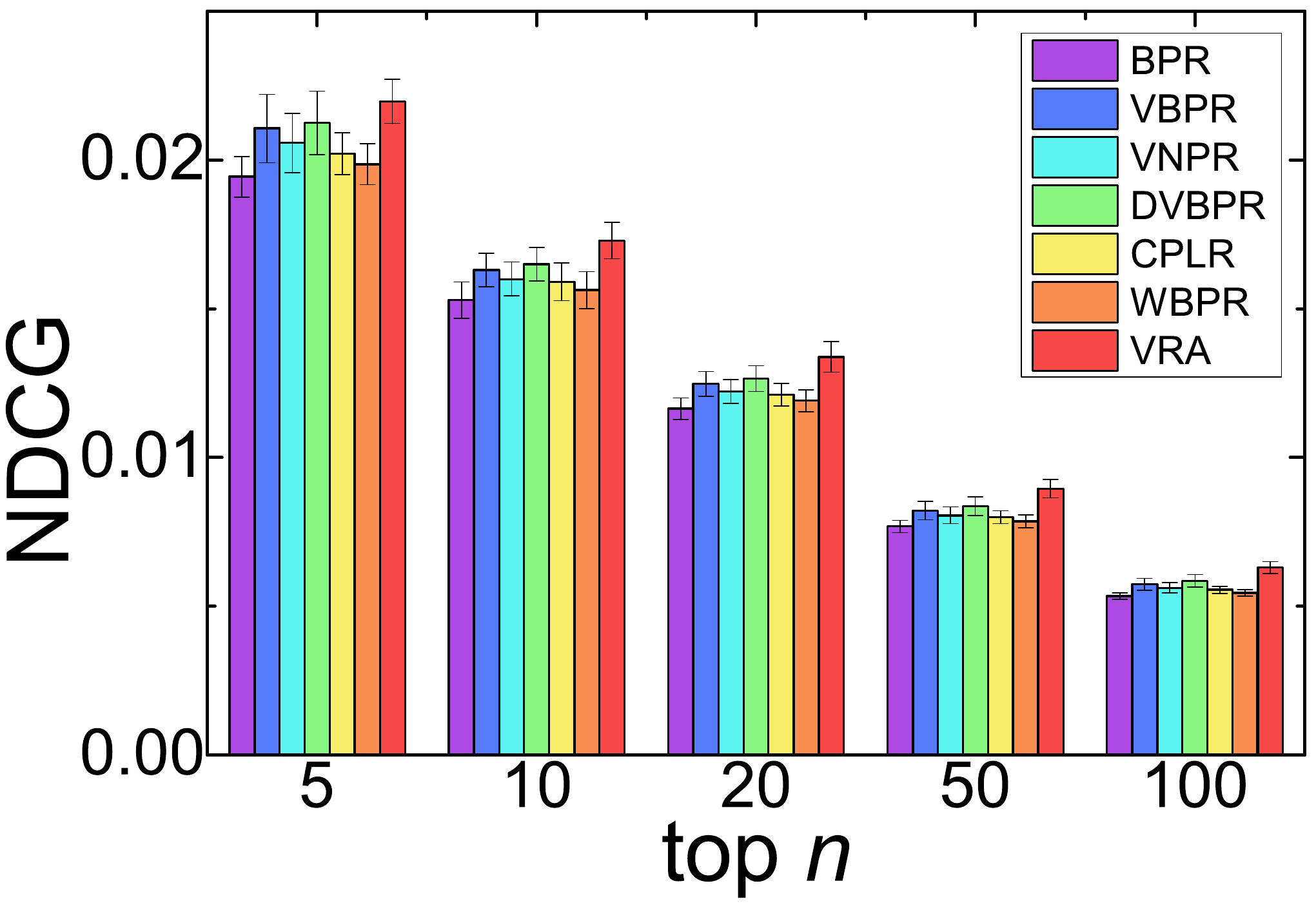}
		\label{fig:women_NDCG}
	}
	\subfigure[\textit{Men}]{
		\includegraphics[scale = 0.25]{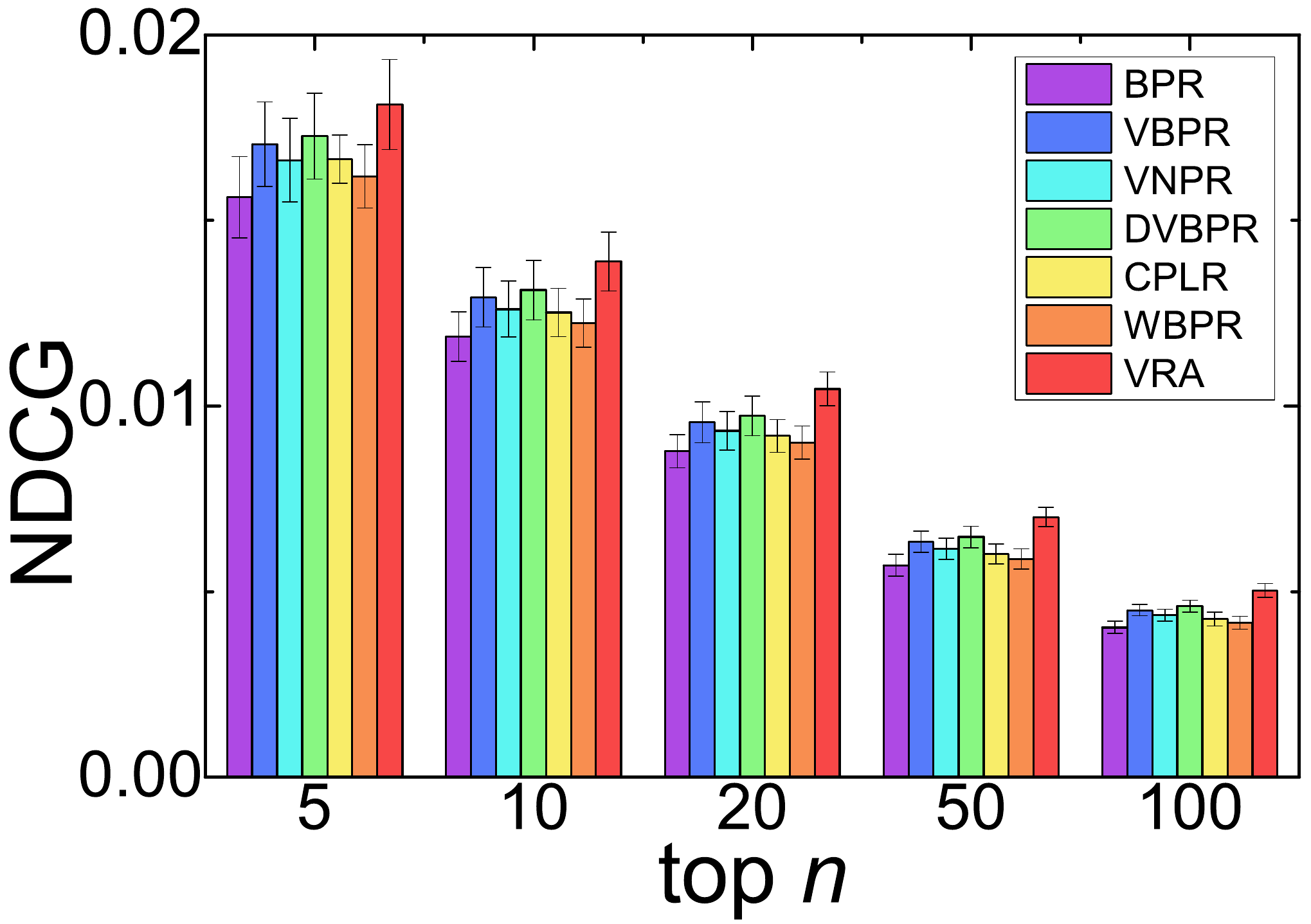}
		\label{fig:men_NDCG}
	}
	\subfigure[\textit{Clothes}]{
		\includegraphics[scale = 0.25]{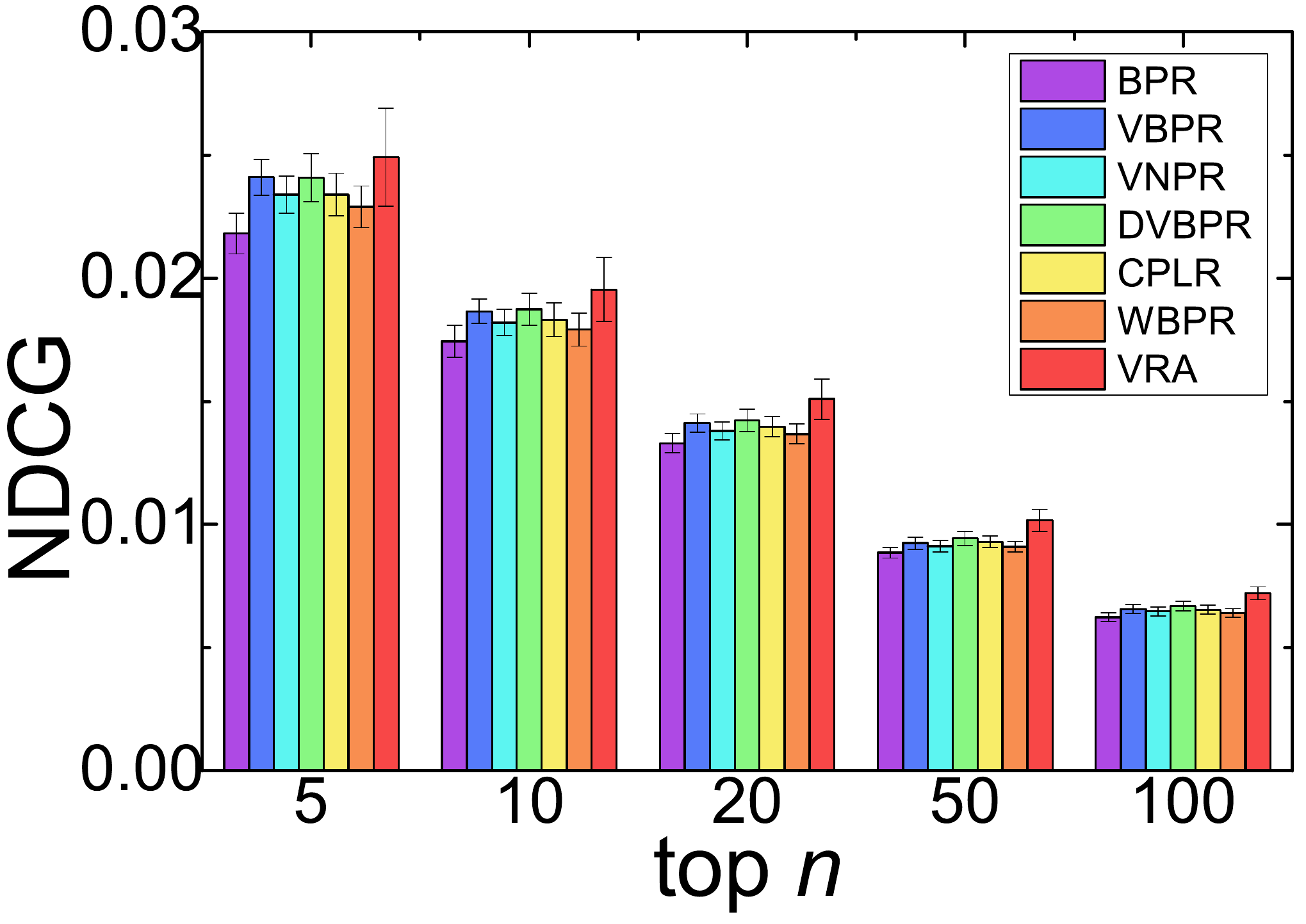}
		\label{fig:clothes_NDCG}
	}
	\subfigure[\textit{Shoes}]{
		\includegraphics[scale = 0.25]{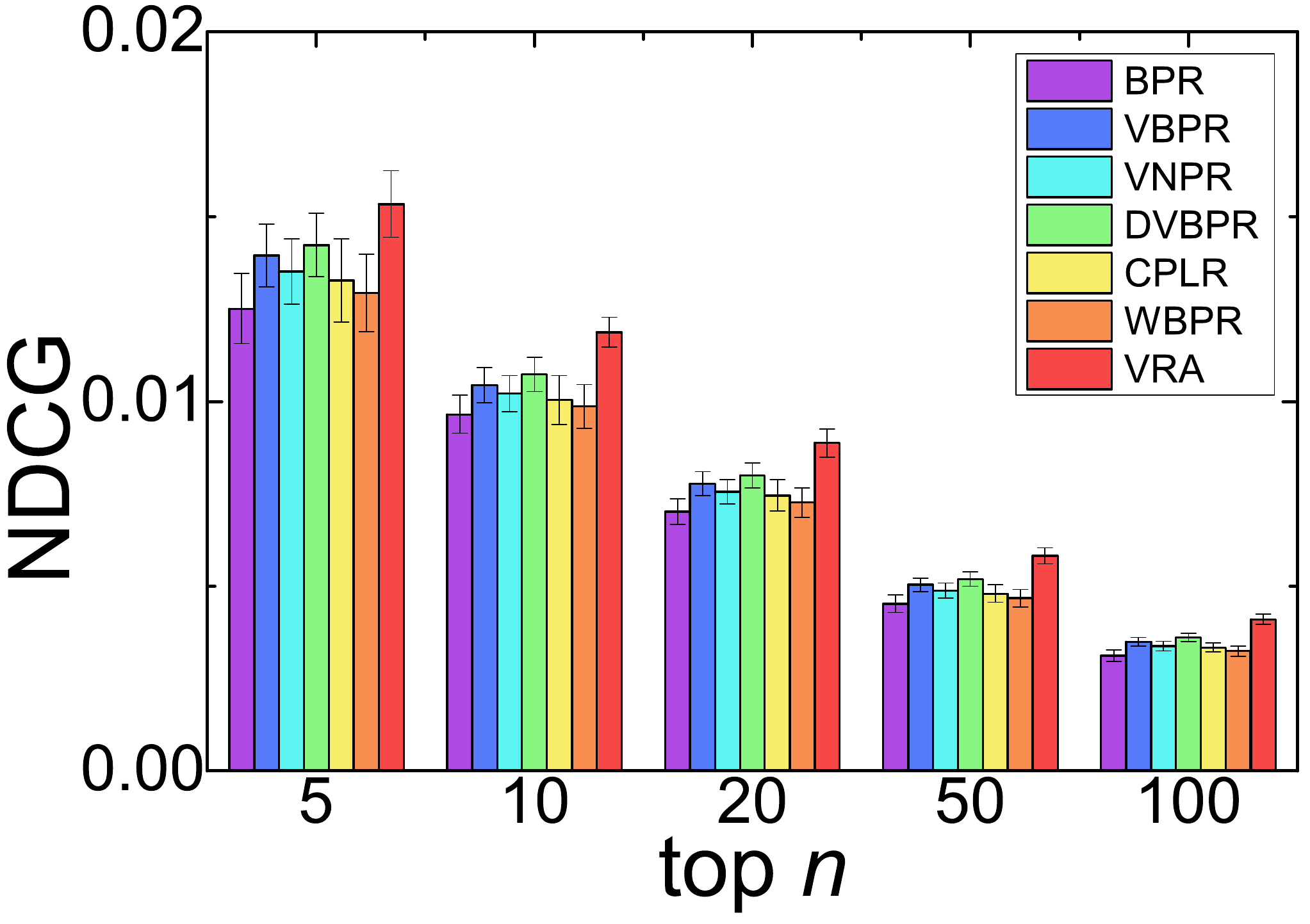}
		\label{fig:shoes_NDCG}
	}
	\subfigure[\textit{Jewelry}]{
		\includegraphics[scale = 0.25]{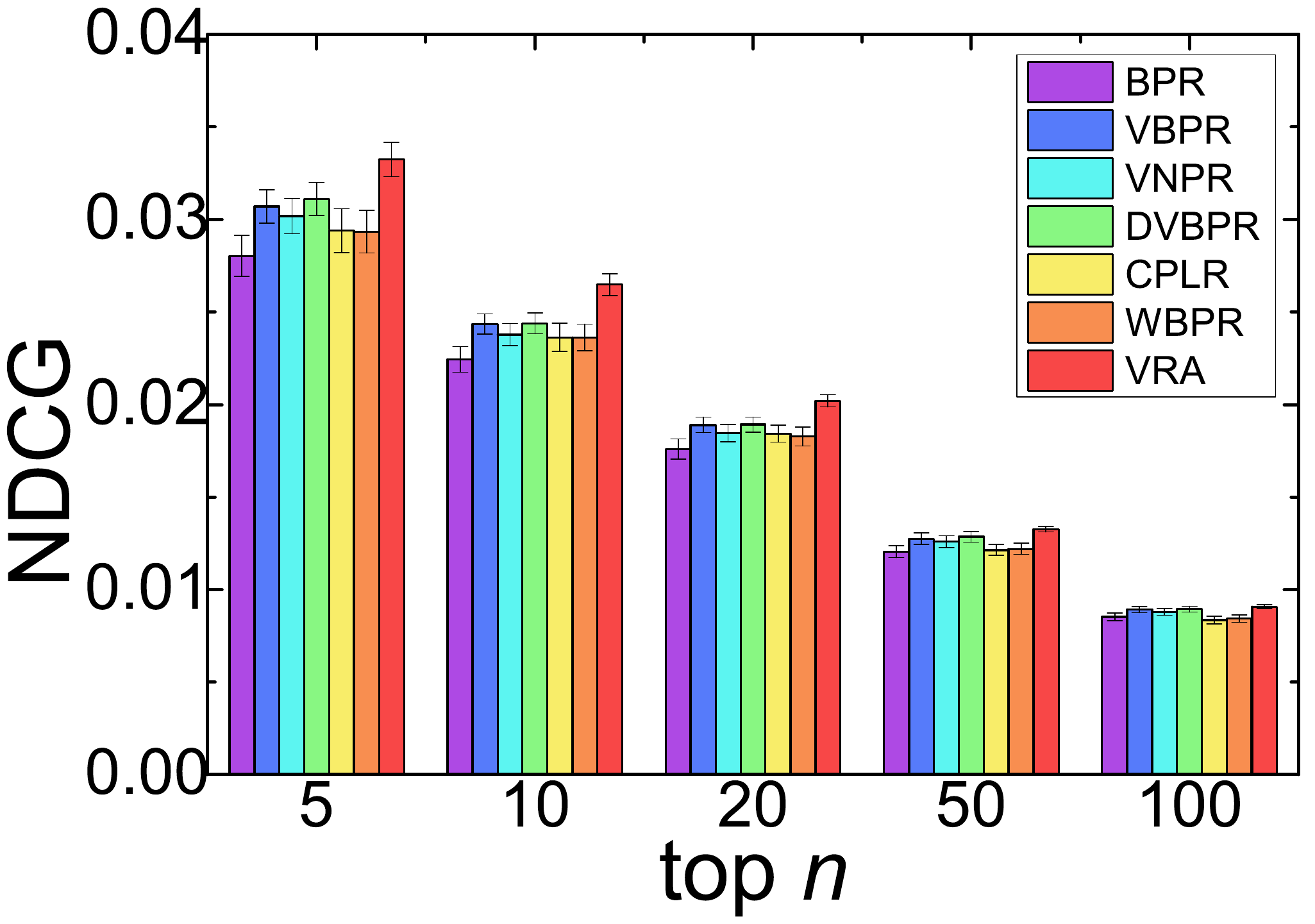}
		\label{fig:jewelry_NDCG}
	}
	\caption{NDCG of different datasets (test set)}
	\label{fig:perf_NDCG}
\end{figure*}

For fair comparison, all models are tuned with the same strategy. We iterate 200 times to train all models. In each iteration, we enumerate all training data on the training set to update the model and and select 1000 samples randomly from the test/validation set to test all models. We record the best performance of each model during this procedure as the evaluation of it. The sampling rate $\rho$ is set as 5 in our experiments. We show the $F_1$-score and NDCG with different $n$ in Figures~\ref{fig:perf_F1} and~\ref{fig:perf_NDCG} respectively. Subfigures (a) to (f) show the performance on \textit{Amazon}, \textit{Men}, \textit{Women}, \textit{Clothes}, \textit{Shoes} and \textit{Jewelry}, respectively. For all datasets and all models, we repeat our experiments 10 times. The bars in Figures~\ref{fig:perf_F1} and~\ref{fig:perf_NDCG} indicate the average performance and the vertical lines on the top of the bars indicate the standard deviation. We can see that the datasets with higher sparsity show lower performance. 

As we can see, BPR performs the worst since it only models user preference based on embeddings. Compared with BPR, advanced baselines including VBPR, VNPR, DVBPR, CPLR, and WBPR use extra information, or advanced sampling strategy, thus outperform it. VBPR, VNPR, DVBPR utilize visual features to model user visual preference and give prediction based both visual features and embeddings. VBPR is the basic visual recommendation model, which simply injects visual features to an MF model. VNPR and DVBPR are deep visually aware models. VNPR inputs user and item embeddings into a deep neural network to achieve better interaction. However, VNPR fails to outperform VBPR in our experiments. DVBPR trains CNN in an end-to-end way to extract fashion aware visual features. As shown in Figures \ref{fig:perf_F1} and \ref{fig:perf_NDCG}, DVBPR outperforms VBPR marginally.

CPLR and WBPR achieve better negative sampling, thus outperform BPR. CPLR utilizes the collaborative information to uncover the potential positive samples, and achieves significant improvement. WBPR weights samples based on item popularity and also performs better than BPR. However, the improvement is marginal since the strategy is simple and rough.

Enhanced by aesthetic features in both modeling and negative sampling aspects, our proposed VRA outperforms all baselines on all datasets. Taking \textit{Jewelry} as example, the proposed VRA model outperforms the best baseline DVBPR about $7.16\%$ on $F_1$-score@10 and $8.64\%$ on NDCG@10.

An interesting observation is that we gain more improvement by the aesthetic feature on \textit{Shoes} and \textit{Clothes} subsets than on \textit{Jewelry} subset. The possible reason is that compared with \textit{Shoes} and \textit{Clothes}, the style of \textit{Jewelry} is relatively simple. For example, the color is almost either silver or golden. In this case, we gain less improvement by modeling the aesthetic features.

\begin{figure}[ht!]
	\centering
	\subfigure[Weighting Parameter]{
		\includegraphics[scale = 0.192]{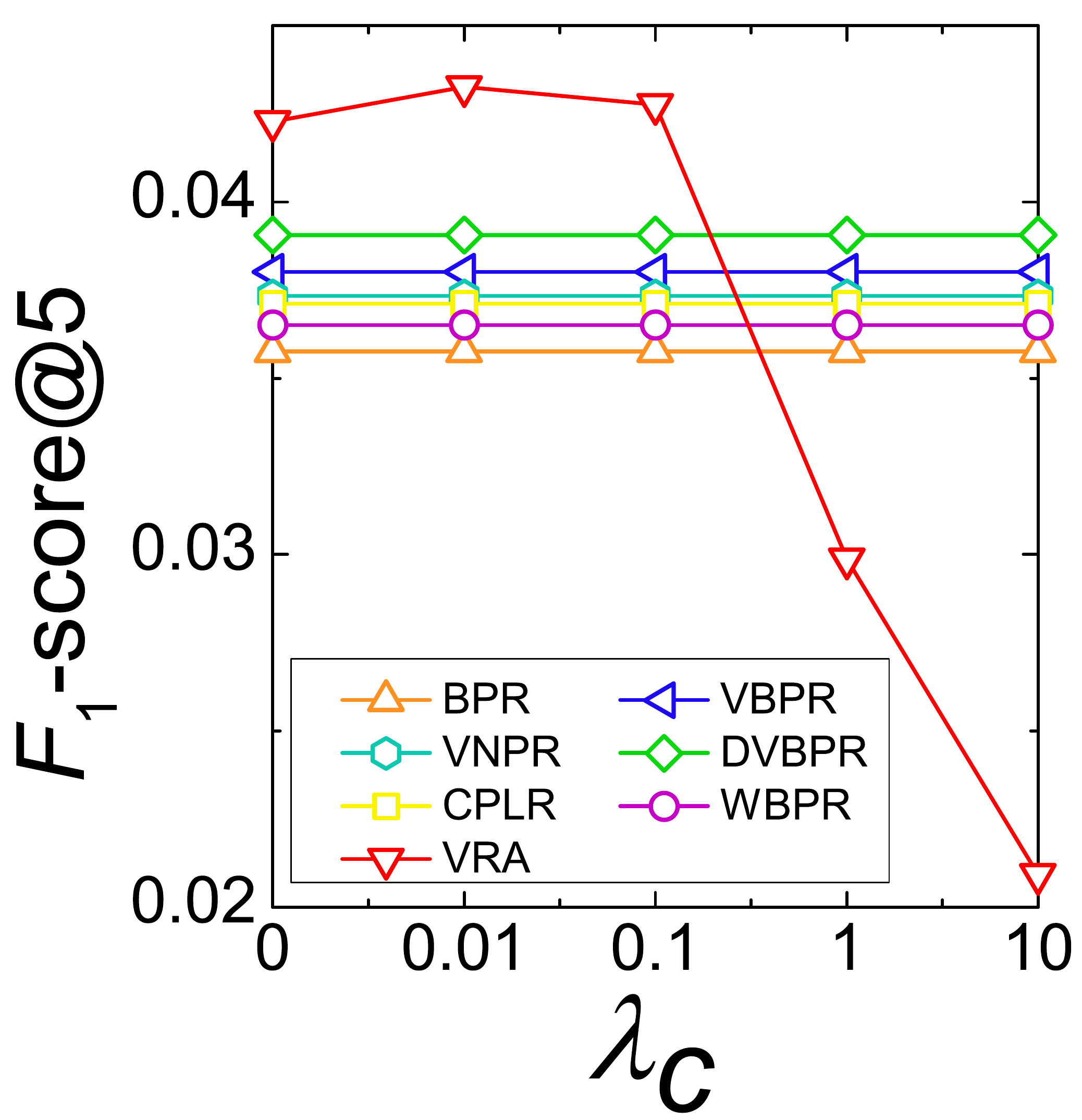}
		\label{fig:lc}
	}
	\hspace{5mm}
	\subfigure[Regularization Coefficient]{
		\includegraphics[scale = 0.3]{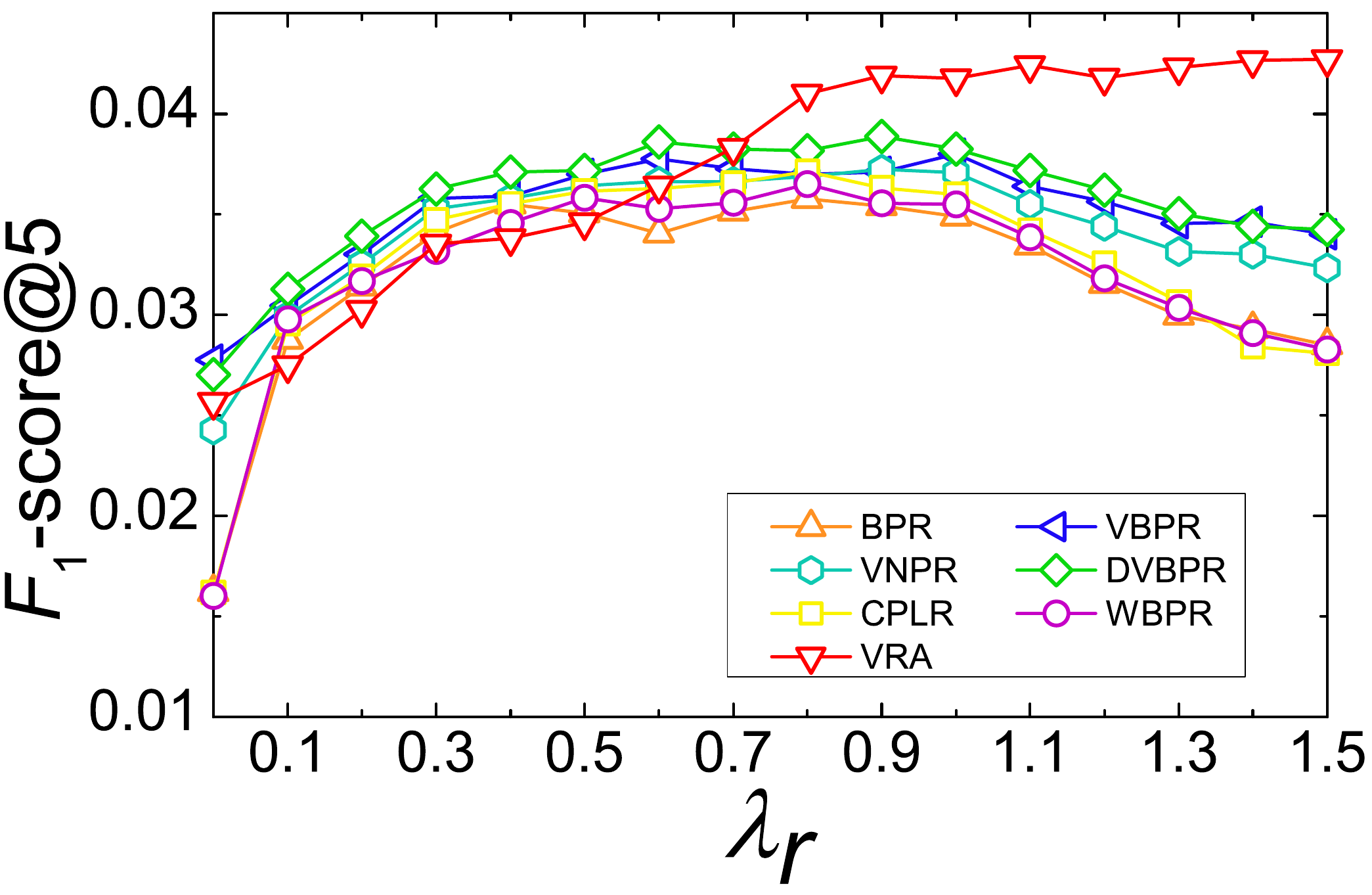}
		\label{fig:lr}
	}
	\caption{Impacts of hyperparameters (\textit{Jewelry}, validation set)}
	\label{fig:f6}
\end{figure}

The sensitivity of $\lambda_c$ and $\lambda_r$ is shown in Figure \ref{fig:f6}. To save space, we only show the result on \textit{Jewelry} set. $\lambda_c$ is a weighting parameter for the coupled matrices (Figure \ref{fig:lc}). When $\lambda_c=0.01$, our model achieves the best performance. The sensitivity of regularization coefficient $\lambda_r$ is shown in Figure \ref{fig:lr}. Our VRA achieves the best performance when $\lambda_r=1.5$ and baselines achieve the best performance when $\lambda_r$ is around 0.9.

\begin{figure}[ht!]
	\centering
	\includegraphics[scale = 0.35]{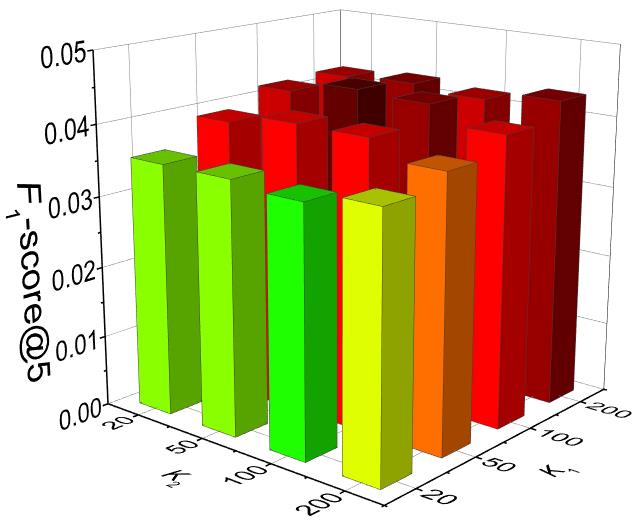}
	\caption{Performance with different length of latent factors (\textit{Jewelry}, validation set).}
	\label{fig:latent_features}
\end{figure}

Figure~\ref{fig:latent_features} shows the performance with different length of latent factors. $K_1$ is the length of latent factors connecting users and items, $K_2$ is the length of latent factors connecting items and time. As Figure~\ref{fig:latent_features} shows, the performance varies with $K_1$ obviously, while not so obviously with $K_2$. It may be because the rank of the user-item matrix ${\bm{{\rm B}}}$ is much higher than that of the time-item matrix ${\bm{{\rm C}}}$, and we need more representation ability to model users' preferences, so our model is more sensitive with $K_1$. 

We can see the rank of \textit{Jewelry} is small: when $K_1=50$, $K_2=20$, the model performs very well and with the increasing of $K_1$ and $K_2$, the performance keeps stable. Considering the ranks of other datasets are higher (such as \textit{All} and \textit{Clothes}) thus larger $K_1$ and $K_2$ are required, we set $K_1=200$ and $K_2=200$ for all datasets. 

\subsection{Necessity of the Aesthetic Features (RQ3)}

In this subsection, we discuss the necessity of the aesthetic features. We combine various widely used features to our basic model and compare the effectiveness of each feature by constructing five models:

\begin{itemize}
	\item{\textbf{VRA\_basic:} This is our basic \textbf{V}isually Aware \textbf{R}e-commendation model without any visual features, which is represented in Subsection~\ref{subsec:dcf}.}
	
	\item{\textbf{VRH:} This is a \textbf{V}isually Aware \textbf{R}ecommendation with Color \textbf{H}istograms. We only inject color histograms to our proposed basic model VRA\_basic.}
	
	\item{\textbf{VRCo:} This is a \textbf{V}isually Aware \textbf{R}ecommendation with \textbf{C}NN Features \textbf{o}nly. We only inject CNN features to VRA\_basic.}
	
	\item{\textbf{VRAo:} This is a \textbf{V}isually Aware \textbf{R}ecommendation with \textbf{A}esthetics Features \textbf{o}nly. We only inject aesthetic features to VRA\_basic.}
	
	\item{\textbf{VRA:} This is our proposed model, utilizing both CNN features and aesthetic features.}
\end{itemize}

\begin{figure}[ht!]
	\centering
	\includegraphics[scale = 0.3]{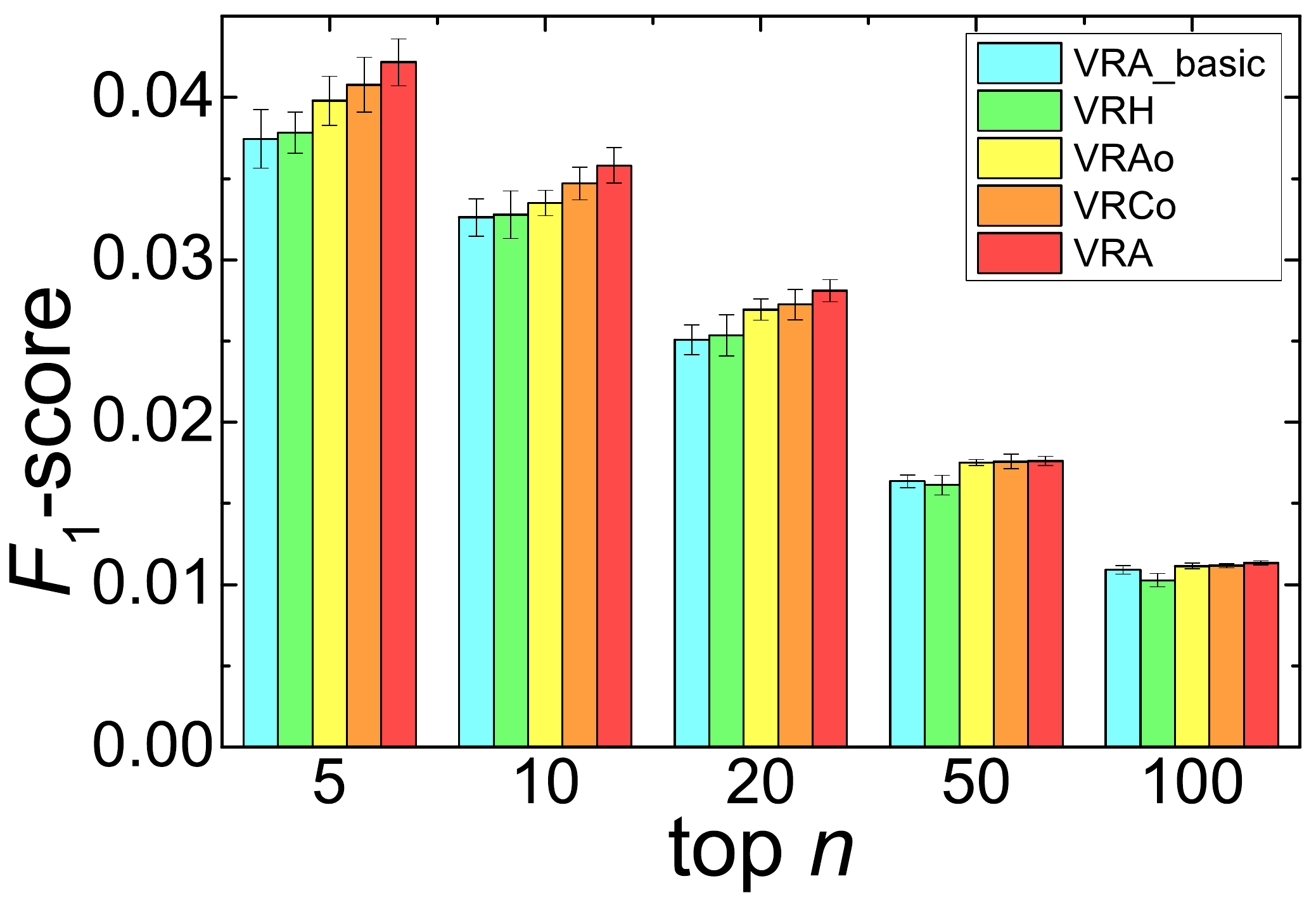}
	\caption{Performance of various features (\textit{Jewelry}, test set)}
	\label{fig:figure6}
\end{figure}

\begin{figure*}[ht!]
	\centering
	\subfigure[]{
		\label{fig:fa} 
		\includegraphics[scale = 0.8]{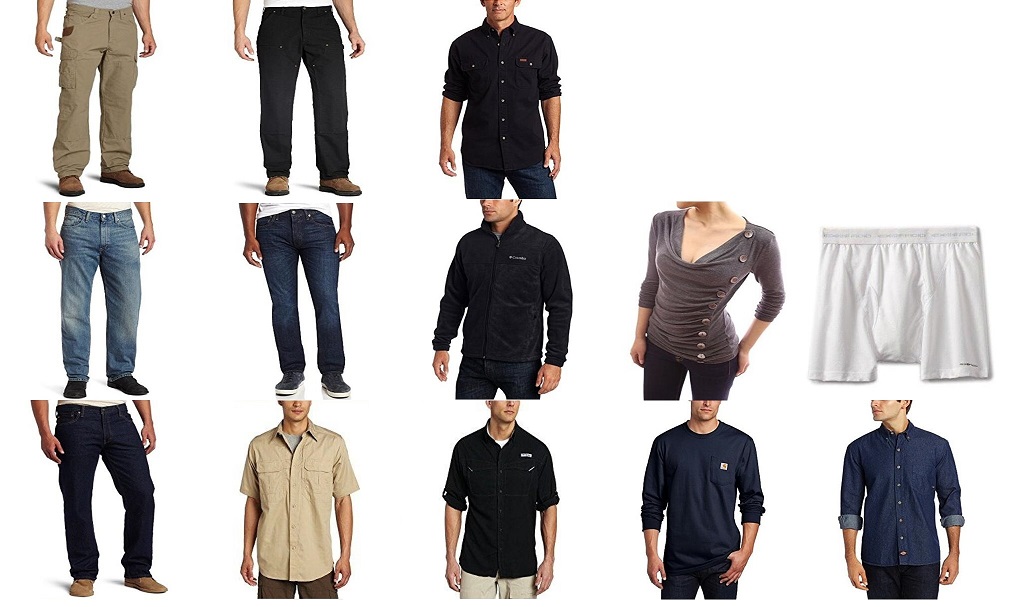}}
	\hspace{5mm}
	\subfigure[]{
		\label{fig:fb} 
		\includegraphics[scale = 0.8]{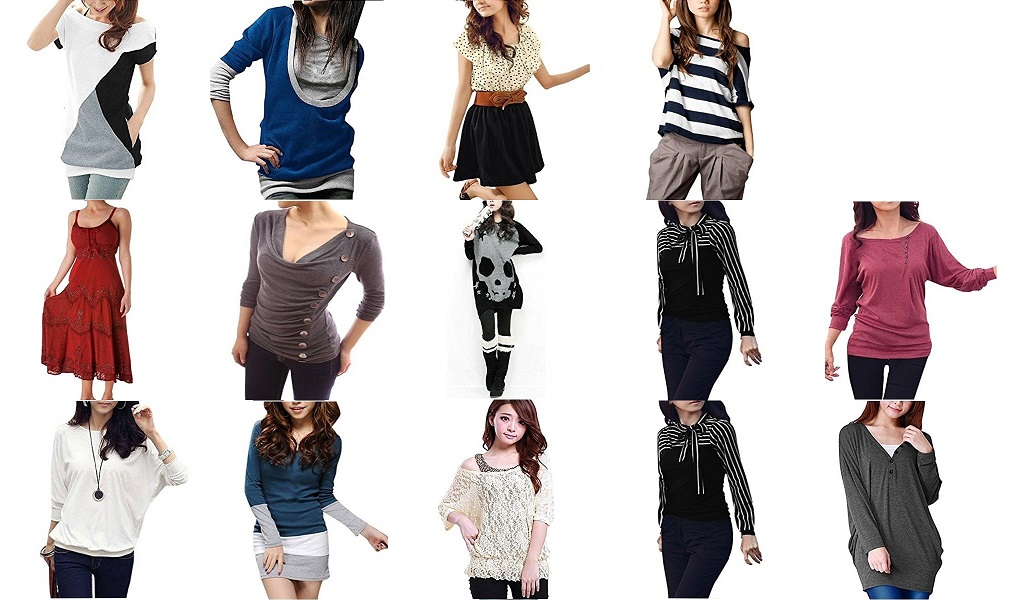}}
	\subfigure[]{
		\label{fig:fc} 
		\includegraphics[scale = 0.8]{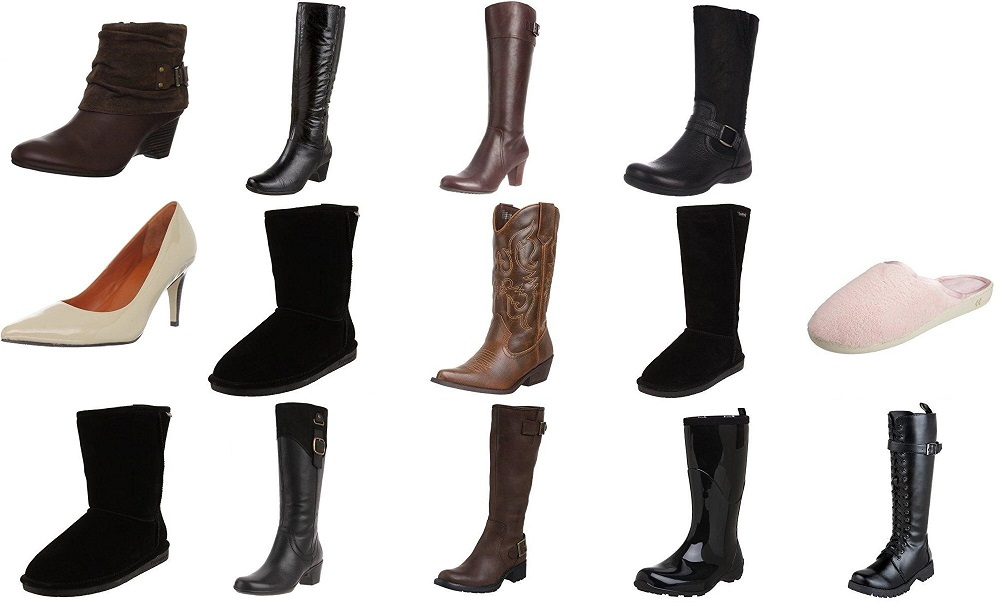}}
	\hspace{5mm}
	\subfigure[]{
		\label{fig:fd} 
		\includegraphics[scale = 0.8]{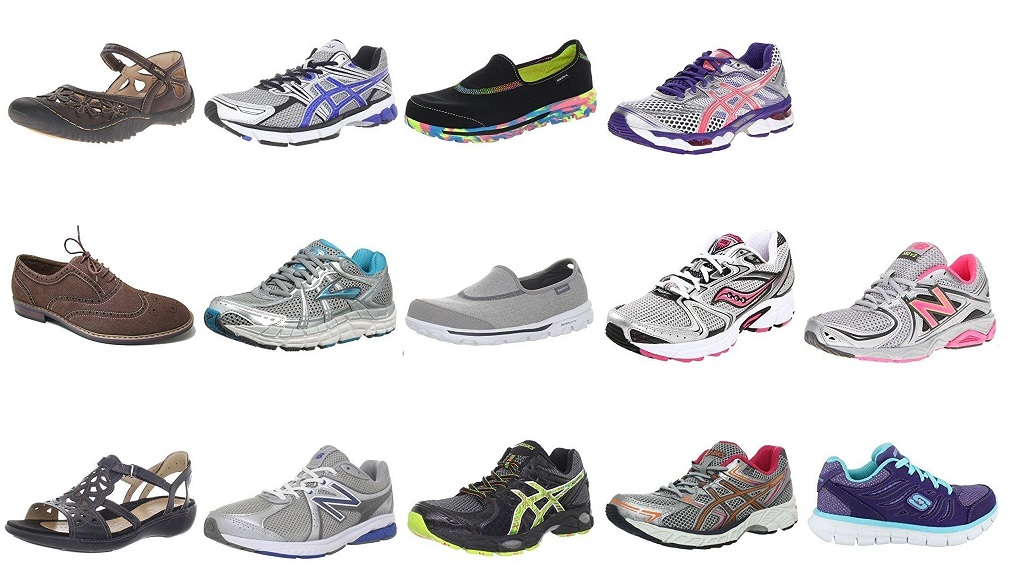}}
	\subfigure[]{
		\label{fig:fe} 
		\includegraphics[scale = 0.8]{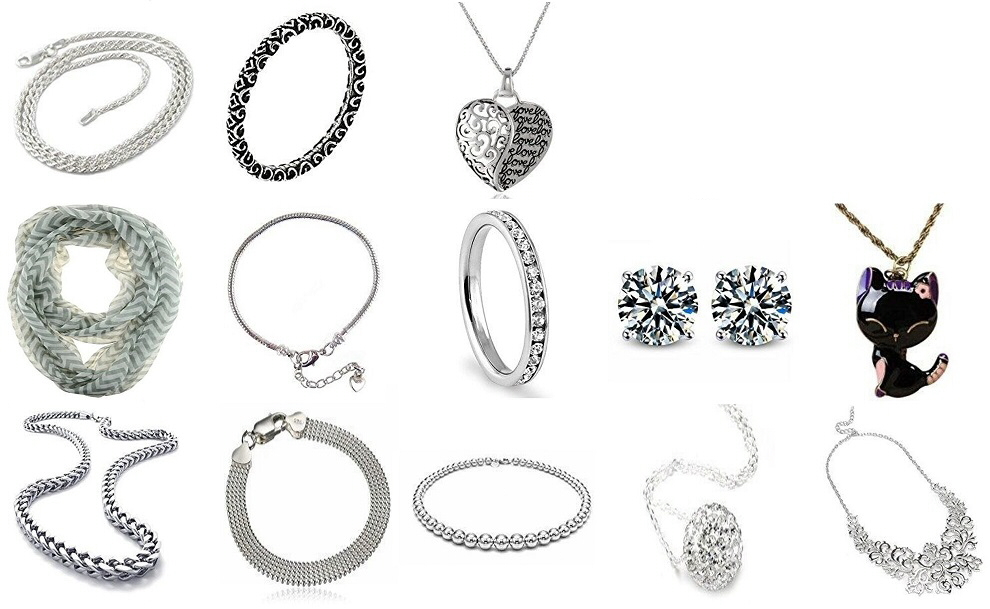}}
	\hspace{5mm}
	\subfigure[]{
		\label{fig:ff} 
		\includegraphics[scale = 0.8]{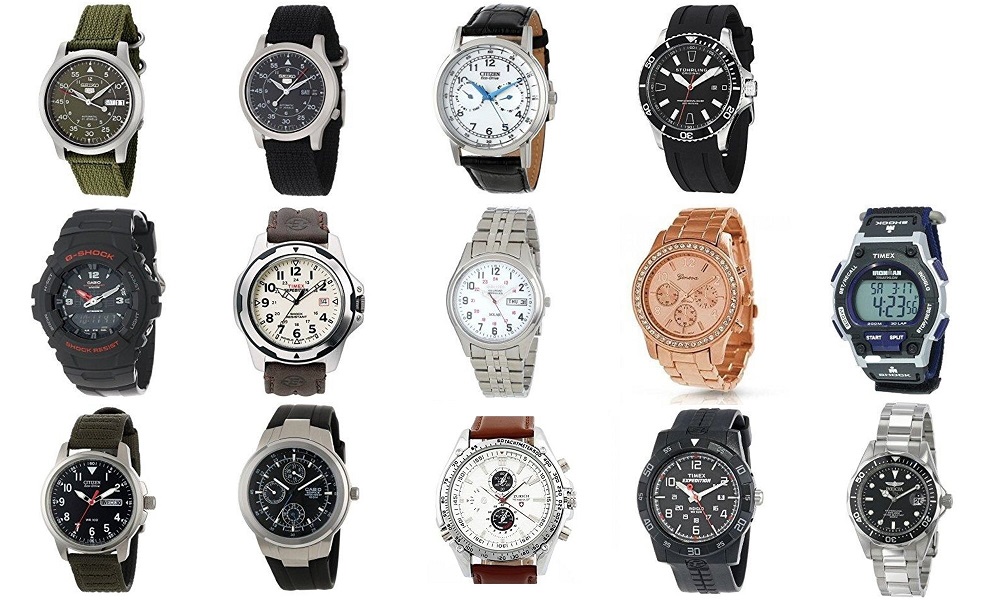}}
	\caption{Items purchased by users and recommended by different models (\textit{Amazon} dataset).}
	\label{fig:compare} 
\end{figure*}

Performances on \textit{Jewelry} dataset are reported in Figure~\ref{fig:figure6}. As we can see, VRA\_basic performs the worst since no visual features are involved to provide the extra information. With the information of color distribution, VRH performs better, though still worse than VRCo and VRAo, because the low-level features are too crude and unilateral, and can provide very limited information about users' aesthetic preferences. VRCo and VRAo show the similar performance because both CNN features and aesthetic features have strong ability to mine user's preferences. Our VRA model, capturing both semantic information and aesthetic information, performs the best on the dataset. 

Figure~\ref{fig:figure6} shows that semantic and aesthetic information mutually enhance each other to a certain extent: they do not perform the best separately, yet they mutually enhance each other and achieve improvement together. Give an intuitive example, if a user wants to purchase a skirt, she needs to tell whether there is a skirt in the image (semantic information) when looking through products, and then she needs to evaluate if the skirt is good-looking and fits her tastes (aesthetic information) to make the final decision.

Several purchased and recommended items on \textit{Amazon} dataset are represented in Figure~\ref{fig:compare}. The items in the first row are purchased by certain user (training data, the number is random). To illustrate the effectiveness of the aesthetic features intuitively, we choose the users with explicit style of preferences and single category of items. The items in the second row and third row are recommended by VRCo and VRA respectively. For these two rows, we choose the five best items from the 50 recommendations to exhibit. Comparing the first and the second row, we can see that leveraging semantic information, VRCo can recommend the congeneric (with the CNN features) and relevant (with tensor factorization) commodities. Although it can recommend the pertinent products, they are usually not in the same style with what the user has purchased. Capturing both aesthetic and semantic information, VRA performs much better. We can see that the items in the third row have more similar style with the training samples than the items in the second row. 

Taking Figure~\ref{fig:fc} as an example, we can see that the user prefers boots, ankle boots, or thigh boots. However, products recommended by VRCo are some different types of women's shoes, like high heels, snow boots, thigh boots, and cotton slippers. Though there is a thigh boot, it is not in line with the user's aesthetics due to the gaudy patterns and stumpy proportion, which rarely appears in her choices. The products recommended by VRA are better. First, almost all recommendations are boots. Then, thigh boots in the third row are in the same style with the training samples, like leather texture, slender proportions, simple design and some design elements of detail like straps and buckles (the second and third ones). Though the last one seems a bit different with the training samples, it is in the uniform style with them intuitively, since they are all designed for young ladies. It is also obvious in Figure~\ref{fig:ff}, we can see that what the user likes are vibrant watches for young men. However, watches in the second row are in pretty different styles, like digital watches for children, luxuriantly-decorated ones for ladies, old-fashioned ones for adults. Evidently, watches in the third row are in similar style with the train samples. They have similar color schemes and design elements, like the intricatel-designed dials, nonmetallic watchbands, small dials, and tachymeters. As we can see, with the aesthetic features and the CNN features complementing each other, VRA performs much better than VRCo.

\subsection{Performance of APLR (RQ4)}

In this subsection, we illustrate the effectiveness of our APLR optimization criterion.

\begin{figure}[ht!]
	\centering
	\includegraphics[scale = 0.4]{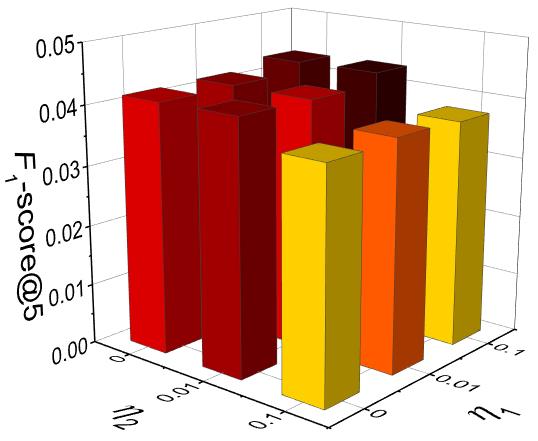}
	\caption{Influence of weighting parameters $\eta_1$ and $\eta_2$ (\textit{Jewelry}, validation set).}
	\label{fig:eta12}
\end{figure}

Figure~\ref{fig:eta12} shows the model tuning with respect to weighting parameters $\eta_1$ and $\eta_2$. We can see that when $\eta_1 = 0.1$ and $\eta_2=0.01$, the model achieves the best performance. When $\eta_1=0$ and $\eta_2=0$, the model becomes VRA\_PLR. As shown in the figure, VRA\_APLR outperforms VRA\_PLR about $4.70\%$ on $F_1$-score.

When $\eta_2$ is fixed, $F_1$-score usually takes the maximum when $\eta_1$ is about 0.1. When $\eta_1$ is fixed, $F_1$-score usually takes the maximum when $\eta_2$ is about 0.01. We come to the conclusion that the preference relationship $(p,q,r)\succ(p,\mathcal{N}_q,r)$ is more important than $(p,\mathcal{N}_q,r)\succ(p,\mathcal{Q}_{pr}^-\setminus\mathcal{N}_q,r)$.

\section{Conclusion}
\label{sec:conclusion}
In this paper, we investigated the usefulness of aesthetic features for personalized recommendation on implicit feedback datasets. We proposed a novel model that incorporates aesthetic features into a tensor factorization model to capture the aesthetic preferences of users at a particular time, {\color{black}{and leveraged visual information and collaborative information to optimize the model}}. Experiments on challenging real-world datasets show that our proposed method dramatically outperforms state-of-the-art models, and succeeds in recommending items that fit users' style. 

For the future work, we are interested in constructing high-order connections among items with spectrum clustering, social networks, etc. instead of only one-order connections, to enhance the pairwise learning. Also, we will establish a large dataset for product aesthetic assessment, and train the networks to extract the aesthetic information better. Lastly, we will investigate the effectiveness of the aesthetic features in the setting of explicit feedback.

\begin{acknowledgements}
This work is supported in part by Beijing Outstanding Young Scientist Program NO. BJJWZYJH0 12019100020098 and National Natural Science Foundation of China (No. 61832017).
\end{acknowledgements}

%
%

\bibliographystyle{spbasic}      

\end{document}